\documentclass[prb,aps,showpacs,twocolumn,amsmath,amssymb,floatfix,superscriptaddress]{revtex4-1}

\usepackage{hyperref}
\usepackage{graphicx}
\usepackage{bm}
\usepackage{dsfont}
\usepackage{xcolor}
\usepackage[utf8]{inputenc}
\usepackage{amssymb}
\usepackage[normalem]{ulem}

\newcommand{\beq}{\begin{equation}}
\newcommand{\eeq}{\end{equation}}
\newcommand{\beqa}{\begin{eqnarray}}
\newcommand{\eeqa}{\end{eqnarray}}

\newcommand{\editE}[1]{\textcolor{black}{#1}}
\newcommand{\editP}[1]{\textcolor{black}{#1}}

\begin{document}

\title{Theory of Caroli--de Gennes--Matricon analogs in full-shell \editE{hybrid} nanowires}
\author{Pablo San-Jose}
\affiliation{Instituto de Ciencia de Materiales de Madrid, Consejo Superior de Investigaciones Cient\'{i}ficas (ICMM-CSIC), E-28049 Madrid, Spain}
\author{Carlos Payá}
\affiliation{Instituto de Ciencia de Materiales de Madrid, Consejo Superior de Investigaciones Cient\'{i}ficas (ICMM-CSIC), E-28049 Madrid, Spain}
\author{C. M. Marcus}
\affiliation{Center for Quantum Devices, Niels Bohr Institute, University of Copenhagen, 2100 Copenhagen, Denmark}
\author{S. Vaitiek\.{e}nas}
\affiliation{Center for Quantum Devices, Niels Bohr Institute, University of Copenhagen, 2100 Copenhagen, Denmark}
\author{Elsa Prada}
\affiliation{Instituto de Ciencia de Materiales de Madrid, Consejo Superior de Investigaciones Cient\'{i}ficas (ICMM-CSIC), E-28049 Madrid, Spain}
 
\date{\today}
 
\begin{abstract}
Full-shell nanowires are hybrid nanostructures consisting of a semiconducting core encapsulated in an epitaxial superconducting shell. When subject to an external magnetic flux, they exhibit the Little-Parks (LP) phenomenon of flux-modulated superconductivity, an effect connected to the physics of Abrikosov vortex lines in type-II superconductors. We show theoretically that full-shell nanowires can host subgap states that are a variant of the Caroli–de Gennes–Matricon (CdGM) states in vortices. These CdGM analogs are shell-induced Van Hove singularities in propagating core subbands. We elucidate their structure, parameter dependence and behavior in tunneling spectroscopy through a series of models of growing complexity. Using microscopic numerical simulations, we show that CdGM analogs exhibit a characteristic skewness towards higher flux values inside non-zero LP lobes resulting from the interplay of three ingredients. First, the orbital coupling to the field shifts the energy of the CdGM analogs proportionally to the flux and to their generalized angular momentum. Second, CdGM analogs coalesce into degeneracy points at flux values for which their corresponding radial wavefunctions are threaded by an integer multiple of the flux quantum. And third, the average radii of all CdGM-analog wavefunctions inside the core are approximately equal for realistic parameters and are controlled by the electrostatic band bending at the core/shell interface. As the average radius moves away from the interface, the degeneracy points shift towards larger fluxes from the center of the LP lobes, causing the skewness. This analysis provides a transparent interpretation of the nanowire spectrum that allows to extract microscopic information by measuring the number and skewness of CdGM analogs. Moreover, it allows to derive an efficient Hamiltonian of the full-shell nanowire in terms of a modified hollow-core model at the average radius.
\end{abstract}

\maketitle

\section{Introduction}

Full-shell nanowires comprised of semiconducting nanowires fully encapsulated in a thin superconducting layer, or shell, have been recently introduced in the context of topological superconductivity~\cite{Vaitiekenas:S20,Penaranda:PRR20}.  These wires offer several advantages for the generation and detection of Majorana bound states (MBSs) as compared to partial-shell ones, where the superconducting coating is limited to some facets of the nanowire~\cite{Mourik:S12,Krogstrup:NM15,Antipov:PRX18,Woods:PRB18,Winkler:PRB19}. In the full-shell case, the trigger of the topological phase transition is the magnetic flux threading the nanowire produced by an external axial magnetic field, whereas in the partial-shell devices following the original proposal \cite{Oreg:PRL10, Lutchyn:PRL10}, it is the Zeeman effect. Partial-shell nanowires, sometimes dubbed Majorana nanowires, have been exhaustively analyzed since 2010, whereas the full-shell variant has only more recently begun being explored \cite{Woods:PRB19,Vaitiekenas:S20,Penaranda:PRR20,Vaitiekenas:PRB20,Kopasov:PSS20,Sabonis:PRL20}.


\begin{figure}
   \centering
   \includegraphics[width=1\columnwidth]{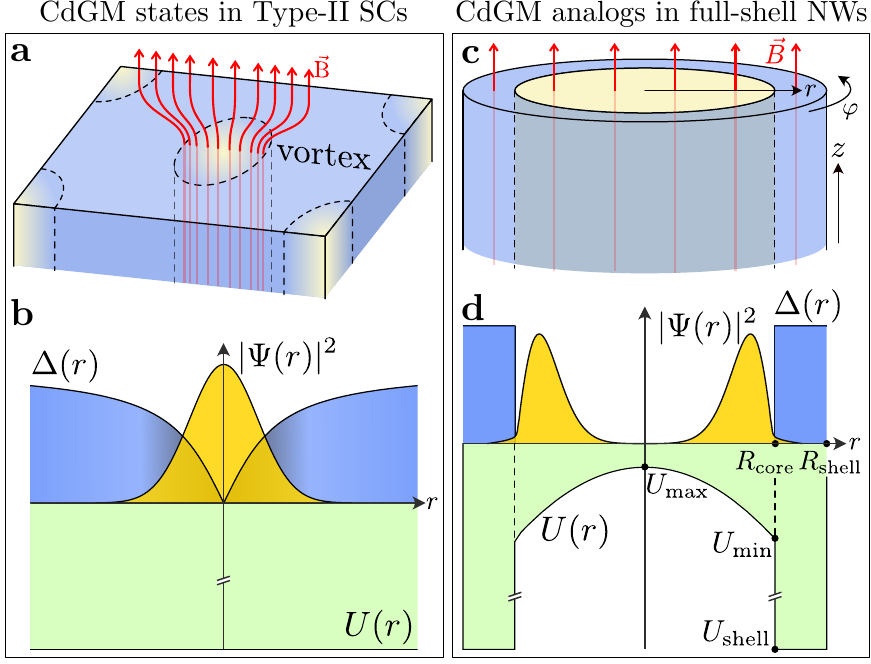}
   \caption{(a) Schematic of Abrikosov vortex lines in a bulk type-II superconductor. The magnetic flux of the external magnetic field $\vec{B}$ inside each vortex is quantized to the superconducting flux quantum $\Phi_0$. (b) Pairing amplitude $\Delta(r)$ (blue), electrostatic potential energy $U(r)$ (green) and lowest-energy CdGM wavefunction density $|\Psi(r)|^2$ (yellow) as a function of radial coordinate $r$ with respect to the vortex center. (c) Schematic of a full-shell hybrid nanowire in a cylindrical approximation. The semiconducting core of radius $R_{\rm{core}}$ (yellow) is fully covered by an $s$-wave superconducting shell of thickness $d_{\rm{shell}}=R_{\rm{shell}}-R_{\rm{core}}$ (blue). The magnetic flux $\Phi$ due to the field $\vec{B}$ threading the wire is not quantized and the pairing amplitude inside the shell is modulated with $\Phi$ following the Little-Parks (LP) effect. (d) Same as (b) but for the full-shell wire. The conduction-band bottom inside the semiconductor exhibits a dome-like radial profile with maximum value at the center, $U_{\rm{max}}$, and minimum value at the superconductor/semiconductor interface, $U_{\rm{min}}$. The electrostatic potential of the metallic shell is $|U_{\rm{shell}}|\gg |U_{\rm{min}}|$.}
   \label{fig:sketch}
\end{figure}


The interest of full-shell hybrid nanowires, however, extends beyond their possible relevance for topological superconductivity. The doubly-connected geometry of the superconducting shell introduces very rich physics~\cite{Vaitiekenas:PRB20,Sabonis:PRL20,Kringhoj:PRL21,Vekris:SR21,Valentini:S21,Escribano:PRB22,Valentini:N22,Ibabe:A22}. In the presence of a magnetic flux $\Phi$ through the section of the hybrid nanowire, the system exhibits the so-called Little-Parks (LP) effect \cite{Little:PRL62, Parks:PR64}. In the LP effect, the flux causes the superconducting phase in the shell to acquire a quantized winding around the nanowire axis. The winding number $n$ is an integer, also known as \textit{fluxoid} number~\cite{London:50,Tinkham:04,De-Gennes:18}, that increases in jumps as $\Phi$ grows continuously. Winding jumps
are accompanied by a repeated suppression and recovery of the superconducting gap, forming LP \textit{lobes} associated with each $n$.
The LP effect has been demonstrated experimentally in various regimes \cite{Liu:S01,Sternfeld:PRL11,Vaitiekenas:PRB20}, and has been shown to be accurately described by theory based on the Ginzburg-Landau formalism \cite{Tinkham:04,De-Gennes:18,Tinkham:PR63,Shah:PRB07,Schwiete:PRL09,Schwiete:PRB10}.

Furthermore, the superconducting boundary condition imposed by the shell gives rise to a special type of fermionic subgap state through a combination of normal and Andreev reflection at the core/shell interface. These states are hybrid-nanowire \textit{analogs} of the celebrated Caroli--de Gennes--Matricon (CdGM) states in Abrikosov vortex lines of type-II superconductors \cite{Caroli:PL64,Brun-Hansen:PLA68,Bardeen:PR69,Tinkham:04}. We call them analogs because both are subgap states within superconducting boundaries, bound to a region with suppressed pairing and threaded by a magnetic flux. However, several important differences exist between them. Some of these were analyzed recently in Refs. \onlinecite{Kopasov:PRB20,Kopasov:PSS20,Kopnin:PRL05}, although these states have remained relatively unexplored.

In bulk type-II superconductors, see Fig. \ref{fig:sketch}(a,b), CdGM states are low energy excitations bound to the center of each vortex core, i.e., to the region of radius $r\lesssim\xi$ (with $\xi$ the bulk superconducting coherence length)~\cite{Caroli:PL64}. Each vortex is threaded by a single flux quantum, which produces a localized suppression of the superconducting order parameter $\Delta(\vec{r})$ at the vortex core $\vec{r}=0$, and a quantized winding of its phase in the polar angle $\varphi$ around the vortex
\beq
\label{Delta}
\Delta(\vec{r}) = \Delta(r)e^{in\varphi}.
\eeq
Here $\Delta(r)=|\Delta(\vec{r})|$ denotes the pairing amplitude, with $\Delta(0)=0$, and the winding is $n=1$.

In full-shell nanowires, on the contrary, the total flux through the core is typically not quantized due to the thinness of the shell, see Fig. \ref{fig:sketch}(c). In this case, both the pairing amplitude and the superconducting gap, dubbed $\Omega$ from here on, are modulated with flux resulting in a series of LP lobes, as mentioned earlier. However, the winding $n$ of the superconducting phase around the shell remains quantized with increasing values in each lobe, making the full-shell wire a multi-fluxoid version of the Abrikosov vortex line \footnote{In this sense, the hybrid wire is more similar to the case of a thin superconducting cylinder or disk that can host a multiquantum giant vortex when its radius is comparable to the coherence length~\cite{Tanaka:PNAS02}.}. Moreover, the confinement of CdGM states inside the type-II superconductor core is dominated by Andreev reflection off the surrounding bulk superconductor, which results in a vanishing group velocity along the vortex line (here the $z$-direction, see Fig. \ref{fig:sketch}). In contrast, in full-shell nanowires the materials of the shell and the core are different, one being a metal with large Fermi energy, and the other a semiconductor with a small Fermi energy.
Even though the shell is epitaxially grown with high quality around the core, the unavoidable velocity mismatch between the two materials produces an abrupt decay of $\Delta(r)$ at the interface, see Fig. \ref{fig:sketch}(d), as well as an enhanced normal reflection for electrons in its interior. As a result, CdGM analogs in the core have a large Fermi velocity along the wire axis, as also noted by Kopasov and Mel’nikov \cite{Kopasov:PRB20}, and form dispersive quasi one-dimensional subbands (as opposed to the non-dispersive CdGM states that are confined inside vortices by strong Andreev reflection).

Another important consequence of the difference in shell and core materials is the band alignment due to their work-function difference, which in InAs produces a significant semiconductor band bending of the Ohmic-type at the interface~\cite{Mikkelsen:PRX18,Schuwalow:AS21}. The conduction-band bending, shown as a dome-like profile of the electrostatic potential $U(r)$ in Fig. \ref{fig:sketch}(d), creates a quantum well at the core/shell interface and thus an accumulation of charge in that region. As a result, the CdGM analogs are typically localized close to the core/shell interface in the radial direction, unlike the CdGM states in Abrikosov vortices~\cite{Berthod:PRL17} [compare $|\Psi(r)|^2$ in Fig. \ref{fig:sketch}(b,d)].

In this work we study the structure and properties of CdGM analogs in realistic full-shell wires, and the information these states can provide about key nanowire aspects through local measurements. We identify the CdGM analogs as Van Hove singularities of the $n$-dependent, quasi-one dimensional, traverse subbands propagating along the axis of the proximitized nanowire core. We use a cylindrical model to describe the hybrid wire, although this approximation is not critical to our findings. Subbands are thus characterized by an angular momentum quantum number, $m_L$, much like the original CdGM sates. They are also characterized by a good radial quantum number (for typical nanowire radii only one or a few radial modes are occupied, in contrast to the many modes occupied in vortices).

We study the energy dispersion of CdGM analogs with magnetic flux inside each $n$-lobe. We compute both local density of states (LDOS) at the end of a semi-infinite wire, and differential conductance ($dI/dV$) through a normal/superconducting junction. The states with $m_L>0$ ($m_L<0$) disperse with a positive (negative) slope versus $\Phi$. This is due to an orbital coupling inside the core of the form $\sim m_L\Phi$. In a simplified hollow-core model for the hybrid wire \cite{Vaitiekenas:S20}, where the semiconductor wavefunction is assumed to be confined to an infinitesimal layer at the core/shell interface, the CdGM analogs disperse with flux symmetrically with respect to the lobe centers, where they coalesce into \textit{degeneracy} points. Meanwhile, realistic solid-core wires, characterized by a potential profile like the one shown in Fig. \ref{fig:sketch}(d), display CdGM analogs with skewed dispersion. Indeed, subgap tunneling $dI/dV$ features in full-shell wires were recently found experimentally to skew towards higher fields within each lobe~\cite{Vaitiekenas:S20}. We find that the spectrum and its \textit{skewness} can be explained by the shift of the degeneracy point towards higher flux $\Phi$ relative to the center of each LP lobe for $n>0$. 

The flux displacement of the degeneracy points is a crucial quantity to understand the subgap spectrum of full-shell nanowires. It is directly related to the spatial wavefunction distribution of the CdGM analogs, or more specifically, to their average radius $R_\mathrm{av}$ inside the core. As we will see, the degeneracy points occur at magnetic fields such that the flux inside $R_\mathrm{av}$ is an integer multiple of the superconducting flux quantum. We perform numerical simulations for Al/InAs full-shell models to show how $R_\mathrm{av}$, and hence the CdGM spectrum within each lobe, is affected by the different materials of the Al shell and the InAs core and the appearance of a charge accumulation layer close to the interface. By direct inspection of the number of subgap states and their skewness, it is possible to determine the effective doping and wave-function distribution in the hybrid wire, which in turn give an approximate measure of the potential profile $U(r)$ inside the semiconductor, see Fig. \ref{fig:sketch}(d) (shaded in green). This provides an indirect but powerful tool to characterize the screened and otherwise inaccessible interior of full-shell devices.
Finally, we discuss a \textit{modified} hollow-core model for the full-shell wire, conveniently tailored to account for the degeneracy-point skewness. We show that it provides very similar results to the full solid-core simulations in the presence of $U(r)$, but at a considerably reduced computational cost.

In most of this work we neglect Zeeman and spin-orbit coupling (SOC) inside the semiconductor since they have a minor effect on the CdGM analogs. Spin is therefore dropped as an inert degree of freedom. The Zeeman effect merely produces small splittings inside small-$n$ LP lobes in the otherwise spin-degenerate subgap spectrum. The SOC, on the other hand, is of course essential for the existence of the topological phase and the emergence of MBSs. As we will show, however, it  leaves the rest of the subgap spectrum practically unaffected.

This paper is organized as follows. In Sec. \ref{sec:LP} we summarize the physics of the LP effect of the shell, deferring the technical details to Apps. \ref{ap:LP} and \ref{ap:SE}. In Sec. \ref{sec:hollow} we characterize the bandstructure, LDOS and quantum numbers of the nanowire modes using the simplified, hollow-core approximation to the full-shell nanowire.
We show how the proximity effect induced by the shell gives rise to Van Hove singularities that become degenerate at special points. In Sec. \ref{sec:tubular} we generalize the model to a finite semiconducting layer thickness, dubbed the tubular-nanowire model, which results in a reduction of the average wavefunction radius of each mode. This leads to a shift of the degeneracy points towards higher fields and the skewness of the CdGM analogs. \editE{Further details on the degeneracy point are given in App. \ref{ap:dpoint}.} In Sec. \ref{sec:solid} \editE{(and App. \ref{ap:Umax})} we connect to the experimentally relevant model of a solid-core nanowire with a finite band-bending electrostatic potential profile in the core. We compute the LDOS at one end of the wire and show how it is related to the tunneling differential conductance of current experiments~\cite{Vaitiekenas:S20,Valentini:S21}. We also develop the \emph{modified} hollow-core description and compare it to the solid core model. In Sec. \ref{sec:SOC} we discuss the effect of including Zeeman and SOC to our models. Finally, in Sec. \ref{sec:discussion} we summarize our main findings and conclude.


\section{The Little-Parks effect of the shell}
\label{sec:LP}

We start by describing the effect of the threading flux $\Phi$ on the superconducting shell alone, i.e., the blue region in Fig. \ref{fig:sketch}(c). Consider a hollow superconducting cylinder along the $\hat{z}$ direction, of thickness $d_\mathrm{shell}$, outer radius $R_\mathrm{shell}$ and inner radius $R_\mathrm{core} = R_\mathrm{shell} - d_\mathrm{shell}$. 
A magnetic field $\vec B = B_z \hat{z}$ is applied along its axis. In the symmetric gauge, the vector potential for $\vec B$ reads $\vec{A}=\frac{1}{2}(\vec{B}\times\vec{r})=(-y, x, 0)B_z/2 = A_\varphi\hat{\varphi}$, where $A_\varphi = B r/2$. Here $r$ is the radial coordinate and $\varphi$ denotes the azimuthal angle around $\hat{z}$. The magnetic field threads a flux through the cylinder, defined as
\beqa
\label{flux}
\Phi &=& \pi R_\mathrm{LP}^2 B_z,\\
R_\mathrm{LP} &=& \frac{R_\mathrm{shell} + R_\mathrm{core}}{2}.\nonumber
\eeqa
Note that $\Phi$ is taken at the mean radius $R_\mathrm{LP}$ of the shell.

In superconductors under magnetic fields, a useful quantity related to $\Phi$ is the \textit{fluxoid} $\Phi'$, which is quantized in units of $\Phi_0=h/2e$, the superconducting flux quantum ~\cite{Deaver:PRL61, Doll:PRL61,Einzel:JLTP11}. This was established by F. London \cite{London:50}, who defined the fluxoid $\Phi'$ as the sum of the magnetic flux $\Phi$ and an extra term involving the superconducting order parameter $\Delta(\vec{r})$, and representing the circulation of persistent supercurrents that arise in response to the magnetic flux. In our hollow cylinder under an axial flux, these supercurrents flow in the $\varphi$ direction, around the cylinder.
If $d_\mathrm{shell}$ is much greater than the London penetration depth~\cite{London:PRSL35,Kittel:96}, $\lambda_L$, the persistent supercurrent term vanishes deep inside the superconductor, and thus the magnetic flux $\Phi$ is also quantized at large distances, $\Phi=\Phi'$. This is the case also in vortices inside bulk superconductors. However, for thin superconducting shells as the ones considered here, the Meissner~\cite{Meissner:N33,Essen:AJP12} expulsion of the magnetic flux is negligible and thus the magnetic field in the superconductor, as well as in its interior, is essentially the same as the applied one (and is hence not quantized). In this case, the second term involving the screening supercurrents oscillates with flux as the fluxoid increases in units of $\Phi_0$, which in turn leads to a modulation of the pairing amplitude $\Delta$, superconducting gap $\Omega$ and critical temperature $T_c$ with a period $\Phi_0$. This is known as the LP effect \cite{Little:PRL62, Parks:PR64,Byers:PRL61,Brenig:PRL61,Groff:PR68}.

\begin{figure}
   \centering
   \includegraphics[width=1\columnwidth]{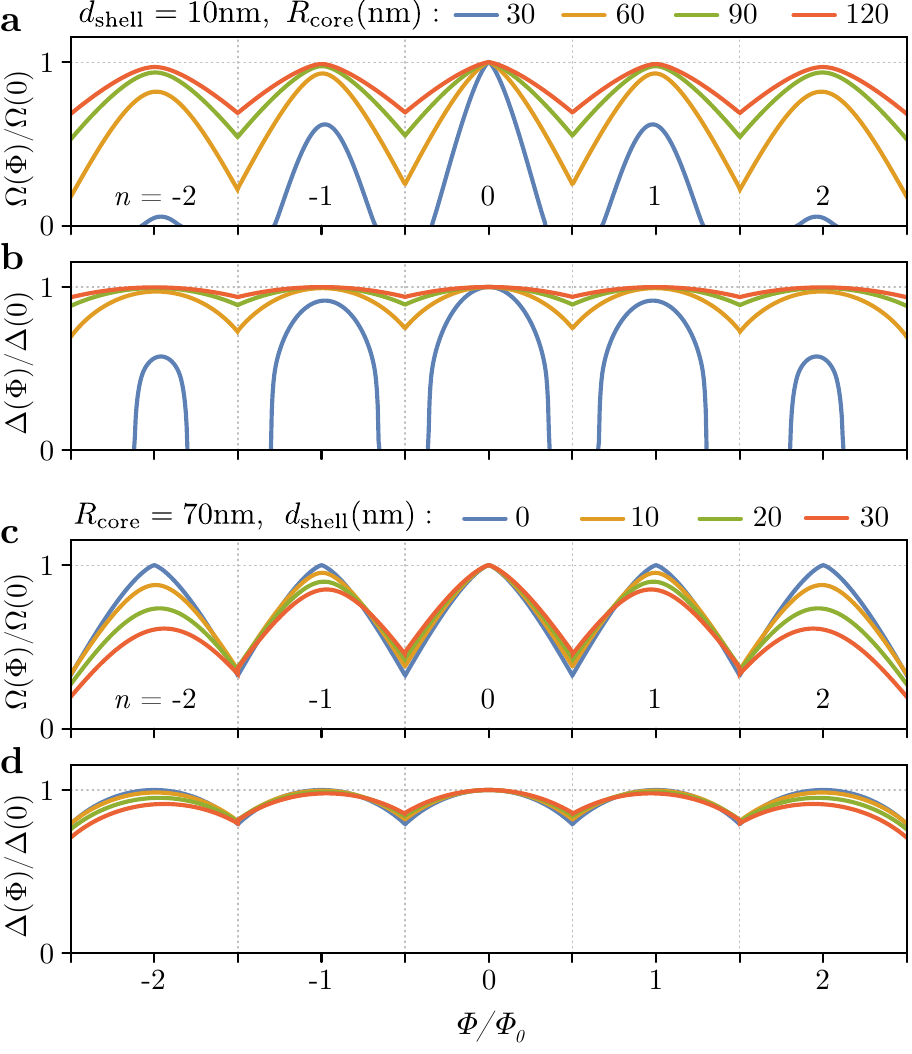}
	\caption{Little-Parks (LP) variation of the shell gap $\Omega$ (a) and pairing amplitude $\Delta$ with applied axial flux $\Phi$ (normalized to the superconducting flux quantum $\Phi_0=h/2e$) of a superconducting cylindrical shell of thickness $d_{\rm{shell}}$, diffusive coherence length $\xi=80$nm and varying internal radii $R_{\rm{core}}$. $n=0,\pm 1, \pm 2$ label the different LP lobes. (c,d) Same as (a,b) but for fixed internal shell radius $R_{\rm{core}}$ and varying $d_{\rm{shell}}$.}
	\label{fig:LP}
\end{figure}

From a complementary point of view, the flux modulation of the superconducting properties is a consequence of the pair-breaking effect of the magnetic field on the Cooper pairs in the superconductor. This pair-breaking effect is minimal at integer values of $\Phi/\Phi_0$, where $\Delta$ and $\Omega$ reach a maximum, and strongest at half-integer values, where they are minimized. Shells with a finite (zero) $\Omega$ at this point are said to be in the non-destructive (destructive) regime, see Fig. \ref{fig:LP}. In the non-destructive case the fluxoid number $n$ and other physical observables undergo an abrupt, first-order transition at lobe boundaries.

Alternatively to the London theory, the quantization of the fluxoid can also be established within the Ginzburg-Landau formalism \cite{Tinkham:PR63,Douglass:PR63} for the complex superconducting order parameter $\Delta(\vec{r}) = \Delta(r,\varphi)$ (we ignore any $z$ dependence). Since this is a single-valued complex quantity, its phase must change by an integer multiple of $2\pi$, $n\in \mathds{Z}$, when completing a closed path around the cylinder, $\varphi\to \varphi + 2\pi$. This winding number is in fact the fluxoid number $n=\Phi'/\Phi_0$. One can thus write $\Delta(\vec{r})$ as in Eq.~\eqref{Delta} for arbitrary $n$.

In this work we are interested in the regime $d_{\rm{shell}}~\ll~\lambda_L$, so that the pairing amplitude is constant, $\Delta(r)=\Delta$. In a ballistic model for the shell, the pairing amplitude $\Delta$ turns out to be equal to the superconducting gap $\Omega$. This is also the case for a time-reversal-symmetric superconductor in the dirty limit according to Anderson's theorem \cite{Tinkham:04,De-Gennes:18}. The shells we consider here are approximated as dirty superconductors, as also done in previous works \cite{Vaitiekenas:S20,Ibabe:A22}. This approximation is reasonable since carriers in experimental shells experience substantial scattering from the typical oxidation layer that develops on the outer surface \cite{Stanescu:PRB17}, domain walls, impurities and even inhomogeneous strains. In the presence of pair-breaking perturbations, like magnetic impurities or (as in our case) magnetic fields in a diffusive superconductor, $\Omega$ is different from $\Delta$. This was originally described by Abrikosov and Gor'kov \cite{Abrikosov:SPU69,Skalski:PR64} whose theory was later applied to the LP effect. The technical details and relevant equations are given in App. \ref{ap:LP}.

The magnetic flux $\Phi$ produces the LP modulation of the shell gap $\Omega(\Phi)$ [Eq. \eqref{LP2}] and pairing amplitude $\Delta(\Phi)$ [Eq. \eqref{LP1}], which exhibit re-emergent lobes centered around integer $n=\Phi/\Phi_0$ as pointed out in the introduction, each of them characterized by a different fluxoid number $n$. The precise $\Omega(\Phi)$ and $\Delta(\Phi)$ profiles depend on the geometric parameters of the shell, and on the superconducting diffusive coherence length $\xi$, see Eq. \eqref{LP3} in App. \ref{ap:LP}. Figure \ref{fig:LP} shows these results for typical nanowire shell parameters.

\section{Hollow-core nanowire}
\label{sec:hollow}

Following Ref. \onlinecite{Vaitiekenas:S20}, we consider a basic model of a cylindrically symmetric full-shell wire that combines (i) the effect of the magnetic flux on the superconducting shell (the LP effect), (ii) the proximity effect on the core subbands with well-defined angular momentum, and (iii) the effect of the magnetic flux on the core subbands.
 
Point (i) is summarized in the preceding section. Regarding (ii), the hybrid wire consists of a semiconducting core with effective mass $m^*$ and radius $R_\mathrm{core}$ covered by a superconducting shell of thickness $d_\mathrm{shell}=R_\mathrm{shell}- R_\mathrm{core}$.
Given a Hamiltonian $H_\mathrm{core}$ for the normal core electrons, we wish to write an effective Hamiltonian $H$ in the presence of the shell by integrating out the shell degrees of freedom. This procedure introduces a self-energy $\Sigma_\mathrm{shell}$ into the Green's function $G(\omega)=\left[\omega - H_\mathrm{core} - \Sigma_\mathrm{shell}(\omega)\right]^{-1}$. The effective Bogoliubov-de Gennes (BdG) Hamiltonian for the system is then defined as $H\equiv\omega-G^{-1}(\omega)=H_{\rm{core}}+\Sigma_\mathrm{shell}(\omega)$, which is in general frequency dependent. Note that we use $\hbar = 1$ throughout, so that $\omega$ has units of energy. 
In the next subsection we define the minimal model that captures also (iii). We call it the hollow-core approximation.


\subsection{Model}



\editE{In the Nambu basis $\Psi=(\psi_\uparrow, \psi_\downarrow, \psi_\downarrow^\dagger, -\psi_\uparrow^\dagger)$, the effective BdG Hamiltonian for the proximitized nanowire then reads}
\beq
\label{hollow3D}
H = \left[\frac{(p_\varphi+eA_\varphi \tau_z)^2 + p_z^2}{2m^*} - \mu\right]\tau_z + \Sigma_\mathrm{shell}(\omega,\varphi),
\eeq
where $p_\varphi = -\frac{1}{r}i\partial_\varphi$, $p_z = -i\partial_z$ are the momentum operators for electrons, $\mu$ is the semiconductor chemical potential, $e>0$ is the unitary charge and $\tau_i$ are Pauli matrices for the electron/hole degree of freedom. The nanowire is subject to a magnetic field as described in Sec. \ref{sec:LP}. Note that both $A_\varphi$ and $p_\varphi$ should be evaluated on the hollow-core surface at $r = R_\mathrm{core}$.

In the expression above, and in general in the rest of this work, we neglect non-local self-energy components (a valid approximation for disordered shells~\cite{Vaitiekenas:S20}) and also any non-uniformity of the self-energy along the wire length, so that $\Sigma_\mathrm{shell}$ depends only on frequency and the angle $\varphi$ around the cylinder axis, $\Sigma_\mathrm{shell}(\omega,\varphi)$. 
As discussed in App. \ref{ap:SE}, the form of $\Sigma_\mathrm{shell}$ for a diffusive shell is expressed in terms of a normal decay rate $\Gamma_{\textrm{N}}$ from the core into the shell and a function $u(\omega)$ given by Eq. \eqref{Eq:u} in App. \ref{ap:LP},
\beqa
\Sigma_\mathrm{shell}(\omega,\varphi)=\Gamma_{\textrm{N}}\frac{\editE{\cos(n\varphi)\tau_x+\sin(n\varphi)\tau_y-u(\omega)\tau_0}}{\sqrt{1-u(\omega)^2}}.
\eeqa
\editE{Note that $u(\omega)$ depends on the flux $\Phi$ and the fluxoid number $n$ through Eq. \eqref{LP3}.}

\subsubsection{Quantum numbers}
\label{subsec:quantnumb}

The hollow-core model in Eq. \eqref{hollow3D} exhibits three symmetries that can be used to classify its eigenstates \cite{Vaitiekenas:S20}. First, $H$ commutes with the electron spin along any direction, so its projection $s_z = \pm \frac{1}{2}$ along $z$ in particular is a good quantum number. Since both values are degenerate we neglect spin until Sec. \ref{sec:SOC}. Second, in the limit of infinite wire, the translation symmetry along $z$ leads to a good $k_z$ quantum number. 
Third, the Hamiltonian exhibits cylindrical symmetry. In the presence of a winding in the pairing $\Delta(r,\varphi)$, the Hamiltonian becomes $\varphi$-dependent, which no longer commutes with the conventional angular momentum $l_z = -i\partial_\varphi$. The generalized angular momentum $L_z = -i\partial_\varphi +\frac{1}{2}n\tau_z$ does commute with $H$, $[L_z, H]=0$, so that the eigenvalues $m_L$ of $L_z$ are good quantum numbers of the eigenstates of $H$. The possible eigenvalues $m_L$ of $L_z$ are \footnote{Note the difference between $m_L$ in this work and $m_J = s_z + m_L = s_z + m_l + n\tau_z/2$ in Ref. \onlinecite{Vaitiekenas:S20}. The reason is the lack of radial SOC in our case, which makes it unnecessary to transform the spin to diagonalize the Hamiltonian.}
\beq
m_L = \left\{\begin{array}{ll}
\mathbb{Z} & \textrm{if $n$ is even} \\
\mathbb{Z} + \frac{1}{2} & \textrm{if $n$ is odd}
\end{array}\right.,
\label{mL}
\eeq
which points to qualitative differences between the spectrum in even and odd LP lobes, as shown in Sec. \ref{sec:LDOS}. The canonical transformation $\mathcal{U} = e^{-i(m_L -\frac{1}{2}n\tau_z)\varphi - ik_z z}$ then reduces $H$ to a $\varphi$-independent $4\times 4$ effective Hamiltonian $\tilde H= \mathcal{U}H\mathcal{U}^\dagger$, where
\beqa
\tilde H &=& \left[\frac{(m_L -\frac{1}{2}n\tau_z+\frac{1}{2}\frac{\Phi}{\Phi_0} \tau_z)^2}{2m^*R_\mathrm{core}^2}+ \frac{k_z^2}{2m^*} - \mu\right]\tau_z \nonumber \\
&&+ \Sigma_\mathrm{shell}(\omega,0).
\label{hollow3Drot}
\eeqa
\editE{The self-energy has here the simpler expression
\beq
\label{Eq:SEmain}
\Sigma_\mathrm{shell}(\omega,0)=\Gamma_{\textrm{N}}\frac{\tau_x-u(\omega)\tau_0}{\sqrt{1-u(\omega)^2}}.
\eeq}
The eigenstates $\tilde\Psi_{m_L,k_z,s_z}$ of $\tilde H$ are related to the original eigenstates $\Psi_{m_L,k_z,s_z}(\varphi, z)$ of $H$ by $\Psi_{m_L,k_z,s_z}(\varphi, z) = \mathcal{U}^\dagger(\varphi, z)\tilde\Psi_{m_L,k_z,s_z}$.

In the absence of proximity effect (that is, for $\Gamma_{\rm{N}} = 0$), the decoupled (normal) core Hamiltonian commutes with the conventional angular momentum $l_z = -i\partial_\varphi$, so that its eigenstates can be classified in terms of the integer eigenvalues of $l_z$, denoted by 
\beq
m_l = \mathbb{Z},
\eeq
together with $k_z$ and $s_z$. In the normal case, then, the canonical transformation that diagonalizes the Hamiltonian is simply $\mathcal{U} = e^{-im_l\varphi - ik_z z}$.

\begin{figure}
   \centering
   \includegraphics[width=1\columnwidth]{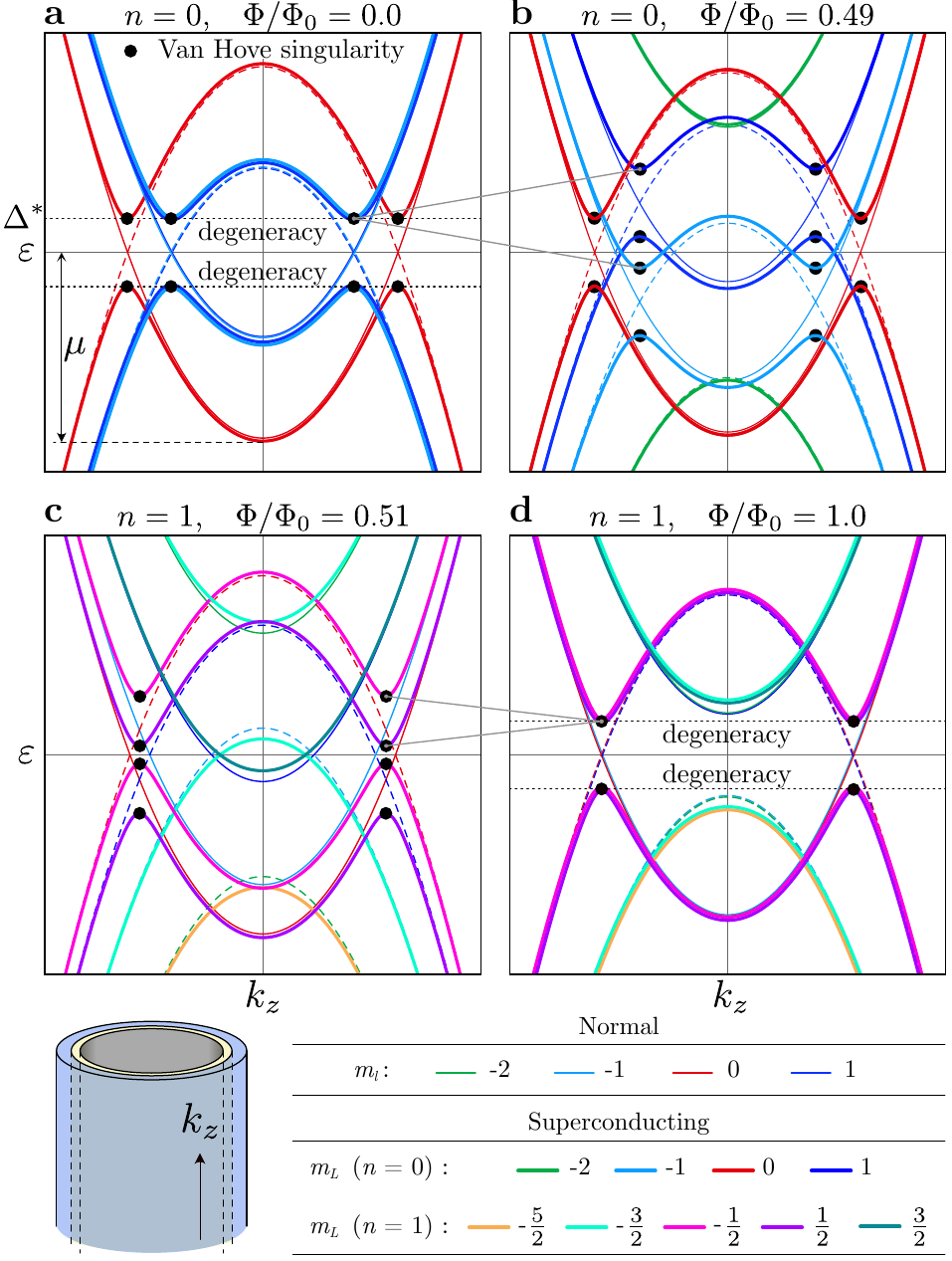}
   \caption{(a) Bogoliubov-de Gennes bandstructure of an infinitely long hollow-core nanowire of radius $R_{\rm{core}}=R_{\rm{shell}}$, as a function of longitudinal momentum $k_z$ at the center of the $n=0$ LP lobe (zero flux). Thin solid (dashed) lines correspond to the normal electron (hole) subbands, with shell/core coupling $\Gamma_{\rm{N}} = 0$, whereas thick lines correspond to the superconducting state, with $\Gamma_{\rm{N}}\neq 0$, \editP{see Eq. \ref{Eq:SEomega0}.} The number of occupied subbands depends on the filling $\mu$. Different colors signal different (generalized) angular momentum number $m_l$ ($m_L$) for the normal (superconducting) subbands. Both $m_l$ and $m_L$ are integers in the $n=0$ lobe, see legend. \editP{The shell-induced superconducting pairing} turns the finite-momentum electron-hole crossings with equal $m_L$ into anticrossings, with Van Hove singularities arising at the edges of the corresponding gaps, see black dots. In the absence of applied flux, $\Phi=0$, all anticrossings are equal in magnitude and centered at zero energy. As a result, all Van Hove singularities are degenerate. (b) Same as (a) but for $\Phi/\Phi_0=0.49$, close to the edge of the $n=0$ lobe. The previously degenerate Van Hove singularities split in energy due to the different dispersion with flux of electron and hole $m_L$ subbands. (c) Same as (b) but for $\Phi$ close to the lower edge of the $n=1$ lobe. The normal bands are very similar to (b), but the anticrossing pattern has changed, as the pairing only couples electron and holes with $m_L$ differing by $n$. The superconducting subband colors represent half-integer $m_L$ quantum numbers. (d) Same as (c) but at the center of the $n=1$ lobe where the hollow core is threaded by one flux quantum. The Van Hove-singularity degeneracies of different subbands are recovered. \label{fig:bands}}
\end{figure}

\subsection{Van Hove singularities and degeneracy points}

We next analyze the Nambu band structure of the hollow-core nanowire model in two different configurations: $\Gamma_{\rm{N}}=0$ (isolated core) and $\Gamma_{\rm{N}}>0$ (proximitized core). \editP{In the latter case we use $\Sigma_{\rm{shell}}(\omega,\varphi)$ evaluated at fixed $\omega = 0$. The $u(\omega)$ solution obeys $u(\omega \to 0) = 0$, so that $\Sigma_{\rm{shell}}(0,\varphi)$ becomes a simple pairing amplitude on the core surface, see Eq. \eqref{Eq:SEomega0}. The resulting band structures}, for four different flux values (two in the $n=0$ lobe and two in the $n=1$ lobe), are shown in Fig.~\ref{fig:bands}. For simplicity we take $R_\mathrm{shell}=R_\mathrm{core}=R_\mathrm{LP}$ in the definition of the flux, Eq. \eqref{flux}. Thin lines are for $\Gamma_{\rm{N}}=0$ Nambu subbands [i.e., normal electron (solid) and hole (dashed) bands]. These are colored according to their corresponding $m_l$ quantum number. Thick lines correspond to finite $\Gamma_{\rm{N}}$, and are labeled by the corresponding $m_L$ quantum number.

The key feature to note in the finite $\Gamma_{\rm{N}}$ bands is the appearance of avoided crossings between normal electron and hole subbands with $m_l$ differing by the fluxoid number $n$. The avoided crossings are due to Andreev reflection at the core/shell interface, and result in Van Hove singularities (black dots) in the LDOS that move in energy with $\Phi$. These Van Hove singularities are the full-shell analogs of CdGM states in type-II superconductors.

The energies of the Van Hove singularities for different $m_L$'s become degenerate at certain values of magnetic flux, see Figs. \ref{fig:bands}(a,d). These special values of $\Phi$ correspond to integer multiples of $\Phi_0$, when the flux $\Phi$ is computed as $\Phi=\pi R_\mathrm{core}^2B_z$, which in the thin-shell limit ($R_\mathrm{LP}=R_\mathrm{core}$) coincides with Eq. \eqref{flux}. Thus, for thin shells and hollow cores, the proximity-induced Van Hove singularities become degenerate at the center of each lobe.

\begin{figure}
   \centering
   \includegraphics[width=1\columnwidth]{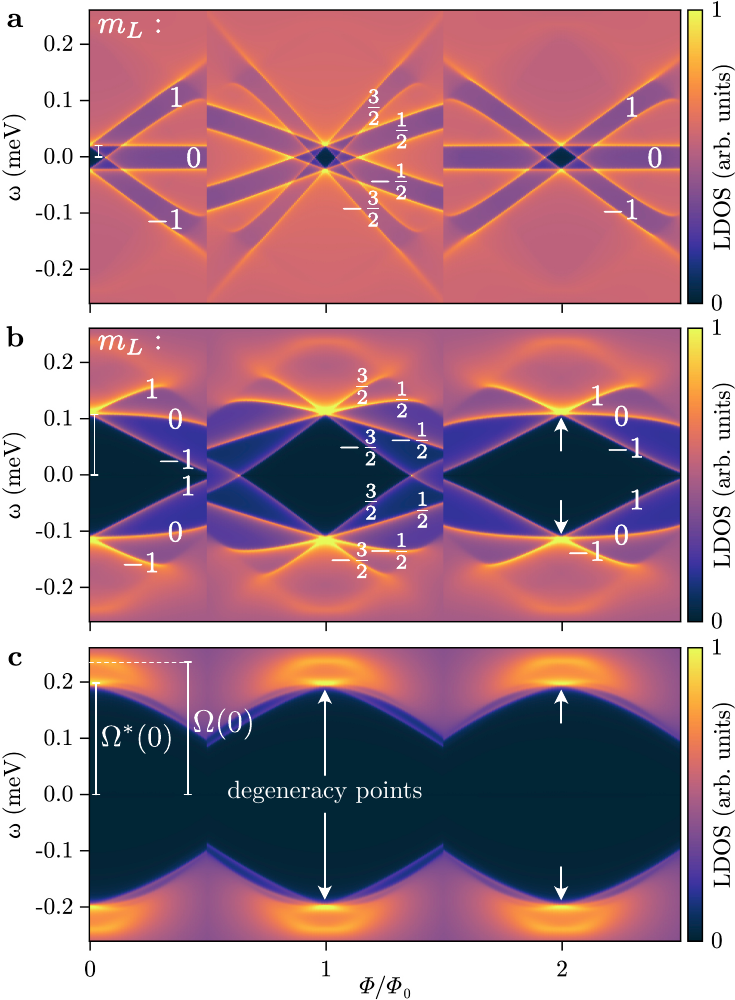}
   \caption{Local density of states (LDOS) at the end of a semi-infinite hollow-core nanowire (in arbitrary units) as a function of energy $\omega$ and applied normalized flux $\Phi/\Phi_0$, displaying half of the $n=0$ lobe, and the full $n=1$ and $n=2$ lobes. CdGM analogs appear as subgap features below the LP shell gap $\Omega(\Phi)$, together with their generalized angular momentum $m_L$. From top to bottom: (a) weak superconductor/semiconductor coupling, $\Gamma_{\rm{N}}=0.1\,\Omega (0)$, (b) intermediate coupling, $\Gamma_{\rm{N}}=0.8\,\Omega (0)$ and (c) strong coupling, $\Gamma_{\rm{N}}=3\,\Omega (0)$. Degeneracy points (where all CdGM analogs cross) happen at the center of each lobe (where the normalized flux $\Phi/\Phi_0$ is an integer), see arrows in (c). The lowest subgap level is dubbed the induced gap $\Omega^*(\Phi)$. Parameters: $R_{\rm{core}}=R_{\rm{shell}}=70$~nm, $\Omega(0)=\Delta(0) = 0.23$~meV, $\mu\approx 1$~meV, $\xi = 70$~nm, $m^*=0.023\,m_e$ and $a_0=5$~nm.}
   \label{fig:degpoint}
\end{figure}

\subsection{Local density of states}
\label{sec:LDOS}

The Van Hove singularities and their degeneracies can also be visualized in terms of the LDOS at the end of a semi-infinite hollow-core nanowire, given by
\beq
\rho(\omega) = -\frac{1}{\pi}\sum_{m_L}\mathrm{Im}\,\mathrm{Tr}\, G^0_{m_L}(\omega).
\eeq
Here, the retarded Green's function $G^0_{m_L}(\omega)$ may be computed 
by discretizing the rotated Hamiltonian $\tilde H$ in a one-dimensional lattice along $z$, with lattice constant $a_0$, and using standard methods of scattering theory \cite{Sanvito:PRB99} to obtain $G^0_{m_L}(\omega)$ in the first unit cell. The superscript $0$ here stands for the first site of the semi-infinite chain. The trace $\mathrm{Tr}$ is taken over the remaining electron/hole degree of freedom. \editP{Here, and in the rest of this work, we restore the full $\omega$ dependence of $\Sigma_{\rm{shell}}$.}

In Fig. \ref{fig:degpoint} we show the calculated LDOS $\rho(\omega)$ as a function of energy $\omega$ and normalized flux $\Phi/\Phi_0$, for different values of core/shell coupling $\Gamma_{\rm{N}}$. This coupling controls the magnitude of the induced gap $\Omega^*(\Phi)$, which is smaller than the gap in the shell $\Omega(\Phi)$. The energy of the coalescing van Hove singularities at zero flux is $\Omega^*(0)$.

In the LDOS simulation of Fig. \ref{fig:degpoint} we have focused on the non-destructive LP regime with $d_\mathrm{shell}\approx 0$, so that the gap edge has the same shape in all lobes. The (spin-degenerate) Van Hove singularities, visible as sharp, flux-dependent subgap features in each lobe, are labeled with their corresponding $m_L$ quantum numbers. Note that each CdGM analog at any given $\Phi$ consists of both the Van Hove singularity itself (seen with a bright orange color in Fig. \ref{fig:degpoint}) and a tail extending above or below it in $\omega$ till the parent gap edge $\pm\Omega(\Phi)$.
It is important to note that only the shell-induced Van Hove singularities (dots in Fig. \ref{fig:bands}) appearing at finite momentum $k_z$ have a good visibility in the LDOS. In contrast, $k_z=0$ Van Hoves, including those already present in the normal bands, are essentially invisible in the LDOS due to the vanishing slope of their electron wave function at the end of the nanowire~\cite{Prada:EPJB04}. \footnote{The absence of visibility in the LDOS of small momentum Van Hove singularities applies also to the $m_J=0$ subband gap closing and reopening at $k_z=0$ that occurs at the topological phase transition in the presence of SOC (not studied here).}


The number of singularities depends on $\mu$ and is different in even and odd lobes. Since $m_L$ is an integer in even lobes (including 0), these contain an odd number of Van Hove pairs. Odd lobes, in contrast, have half-integer $m_L$ [see Eq. \eqref{mL}], so they contain an even number of Van Hove pairs, see Fig. \ref{fig:degpoint}.

We saw in the bandstructures of Fig. \ref{fig:bands} how Van Hove singularities become degenerate  at the center of each lobe. In the LDOS this is visible as coalescing singularities, forming a characteristic fountain-like pattern around degeneracy points, and symmetrically around $\omega=0$. The slope with which the singularities disperse with flux away from the degeneracy points is proportional to $m_L$. This is ultimately due to the orbital coupling term
\beq
\frac{1}{2m^*R_\mathrm{LP}^2}\frac{\Phi}{\Phi_0}m_L\tau_z=\omega_{\mathcal{L}} m_L\tau_z,
\eeq
that is present in Eq. \eqref{hollow3Drot} after expanding the square. Here, $\omega_{\mathcal{L}}=eB_z/2m^*$ is the Larmor frequency~\cite{Brun-Hansen:PLA68}. The slope is not constant, however, as the spectral density at the band edge in $\Sigma_\mathrm{shell}$ repels the CdGM analogs as they approach $\Omega(\Phi)$.

\section{Tubular-core nanowire}
\label{sec:tubular}

\begin{figure*}
   \centering
   \includegraphics[width=\textwidth]{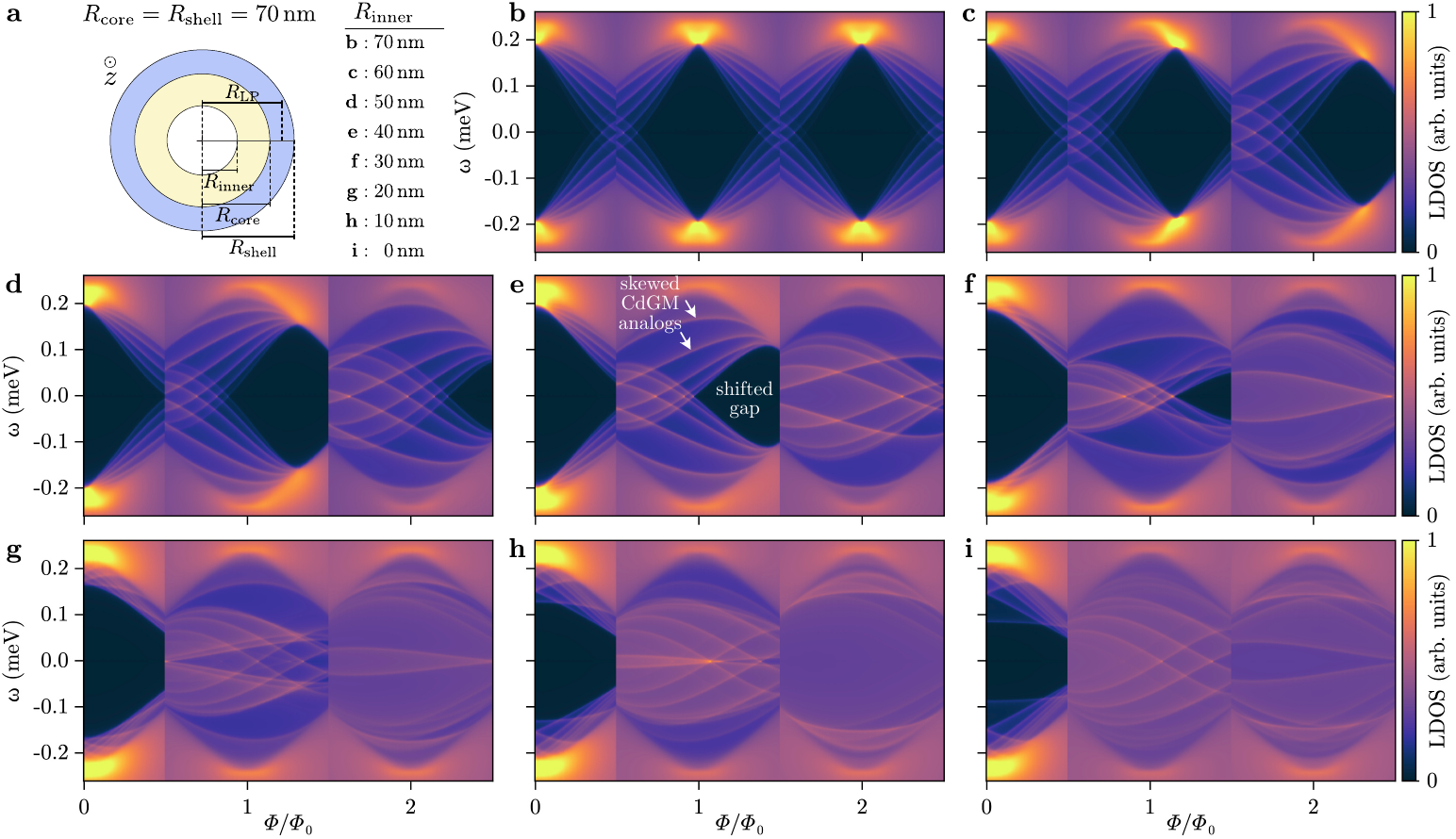}
   \caption{(a) Schematics of the full-shell nanowire cross section for a tubular semiconducting core with inner $R_{\rm{inner}}$ and outer $R_{\rm{core}}$ radii. The outer shell radius is $R_{\rm{shell}}$. (b-i) LDOS in arbitrary units versus energy $\omega$ and normalized flux $\Phi/\Phi_0$ for increasing tubular-core thickness $d_{\rm{core}}=R_{\rm{core}}-R_{\rm{inner}}$, from the hollow-core approximation in (b) to the solid-core case in (i). The corresponding values of $R_{\rm{inner}}$ are displayed in (a). As the core thickness increases, the degeneracy points shift to larger flux within each $n\neq 0$ LP lobe, skewing the CdGM analogs and shifting the gap $\Omega^*$ below them. The electrostatic potential inside the semiconductor is uniform, and is adjusted in each panel to yield 13 (spin-degenerate) $m_L$ subbands occupied at zero flux. The superconductor/semiconductor coupling $\Gamma_{\rm{N}}$ is adjusted to have $\Omega^*(0)\approx 0.2$~meV. Other parameters: $a_0=5$nm, $R_{\rm{core}}=R_{\rm{shell}}=70$~nm, $\Omega(0)=0.23$~meV, $\xi=70$nm.}
   \label{fig:tubular}
\end{figure*}

To continue building towards the more realistic nanowire model discussed in Sec. \ref{sec:solid}, we next generalize the hollow-core model by giving a finite thickness $d_\mathrm{core} = R_\mathrm{core}-R_\mathrm{inner}$ to the semiconductor, so that it spans a finite range of radii $r\in[R_\mathrm{inner}, R_\mathrm{core}]$, see Fig. \ref{fig:tubular}(a), while keeping the potential in the core $r$-independent. We dub this the tubular-core nanowire model.

\subsection{Model}

The tubular-core generalization introduces radial kinetic energy into the model, and consequently radially quantized modes. The corresponding effective BdG Hamiltonian reads
\beq
\label{tubular}
H = \left[\frac{(p_\varphi+eA_\varphi(r) \tau_z)^2 + p_r^2 + p_z^2}{2m^*} - \mu\right]\tau_z + \Sigma_\mathrm{shell}(\omega,\varphi),
\eeq
where $p_r^2 = -\frac{1}{r}\partial_r(r\partial_r)$. Note that we have also restored the radial dependence of $A_\varphi(r) = B_z r/2$. The same canonical transformation $\mathcal{U}$ as for Eq. \eqref{hollow3Drot} reduces the above to
\beqa
\label{tubularrot}
\tilde H &=& \left[\frac{(m_L -\frac{1}{2}n\tau_z+\frac{1}{2}\frac{\Phi}{ \Phi_0} \frac{r^2}{R_{\rm{LP}}^2} \tau_z)^2}{2m^*r^2}+ \frac{k_z^2+p_r^2}{2m^*} - \mu\right]\tau_z \nonumber\\
&&+ \Sigma_\mathrm{shell}(\omega,0).
\eeqa
To find the eigenstates $\tilde \Psi(r)$ of $\tilde H$ we follow the DLL-FDM scheme of Ref. \onlinecite{Arsoski:CPC15}. We first discretize the radial coordinate $r$ with a lattice spacing $a_0$, replacing derivatives with finite differences in the differential eigenvalue equation $\tilde H\psi(r)=\varepsilon\tilde\Psi(r)$. We then absorb the Jacobian $J=r$ of the cylindrical coordinates into modified discrete eigenstates $F(r_i) = \tilde\Psi(r_i)\sqrt{r_i}$ and into the corresponding Hamiltonian $H'= r^{1/2}\tilde H r^{-1/2}$. With this we arrive at a discrete eigenvalue problem $\sum_{i'}{H'}_{ii'} F(r_{i'}) = \varepsilon F(r_i)$ with a Hermitian Hamiltonian matrix ${H'}_{ii'}$, whose discrete eigenstates are, by virtue of their definition, trivially orthonormal without $J$, $\sum_iF^*_\alpha(r_i)F_\beta(r_i)=\delta_{\alpha\beta}$. The kinetic energy  $\tau_z p_r^2/m$ in $H'$ transforms, in the discrete ${H'}_{ii'}$, into an onsite term $o_i=2t_0\tau_z$ plus a radial hopping $t_{ii'} = -t_0\tau_z r/\sqrt{r_ir_{i'}}$ between the nearest neighbors, where $t_0=1/(2m^*a_0^2)$. Note that the $r/\sqrt{r_ir_{i'}}$ factor directly stems from the cylindrical Jacobian, but does not break the symmetry $t_{ii'}=t_{i'i}$. Also, when applying the above DLL-FDM scheme to systems including the origin $r=0$, the correct boundary condition must be implemented there. This is done by excluding the $r=0$ site and multiplying $o_i$ at the $r=a_0$ site by 3/4. \cite{Arsoski:CPC15}

\begin{figure*}
   \centering
   \includegraphics[width=\textwidth]{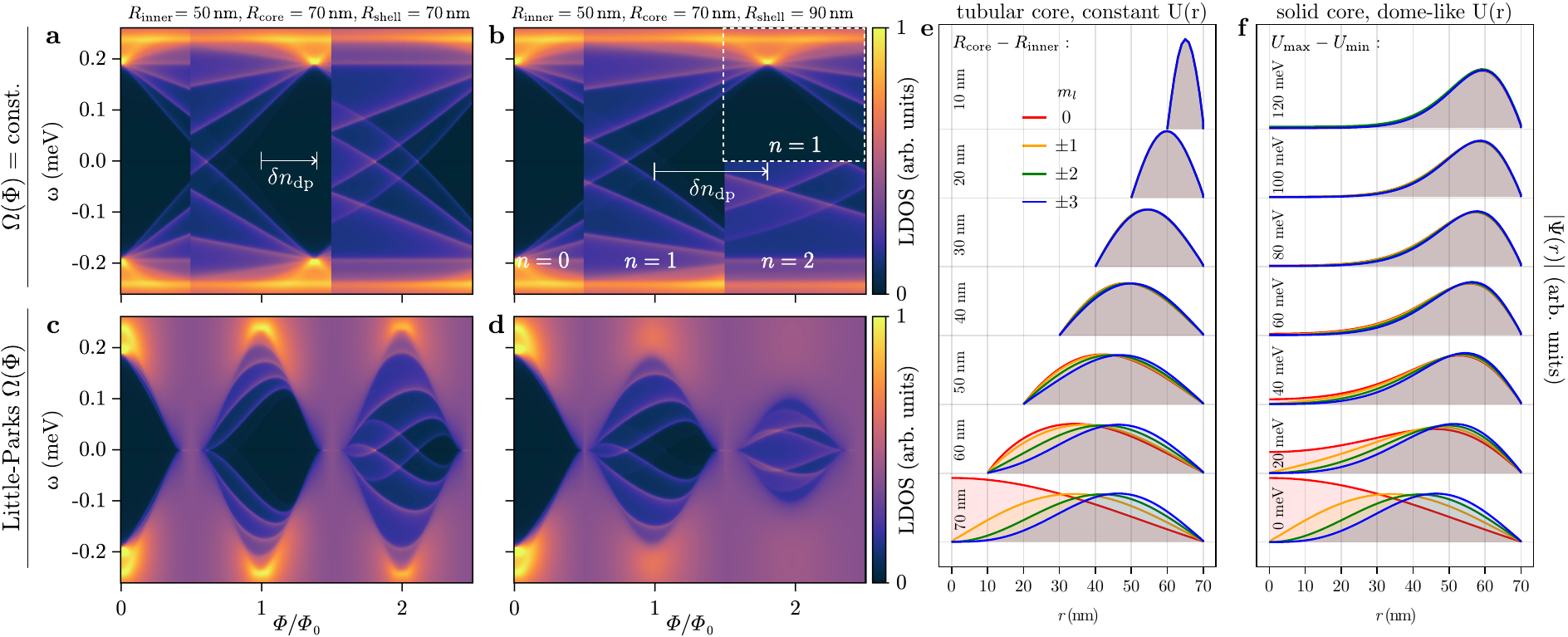}
   \caption{(a) LDOS in arbitrary units as a function of energy $\omega$ and normalized flux $\Phi/\Phi_0$ for a full-shell tubular-core nanowire with $R_{\rm{inner}}=50$~nm and $R_{\rm{core}}=R_{\rm{shell}}=70$~nm, $a_0=5$~nm and core/shell coupling $\Gamma_{\rm{N}}=250\,\Omega(0)$, which yields $\Omega^*(0)\approx 0.2$~meV. The LP gap is forced to be flux-independent ($\xi\to 0$). The degeneracy point in the $n = 1$ lobe shifts to the right by a quantity $\delta n_{\rm{dp}}$ with respect to the $n=1$ lobe center. (b) Same as (a) but for $R_{\rm{shell}}=90$~nm, which shifts the degeneracy point even further. The positive-energy part of the $n = 2$ lobe, enclosed by a white dashed box, has been artificially forced to maintain an $n=1$ fluxoid to reveal the position of the metastable $n=1$ degeneracy point. (c,d) Same as (a,b) but including the LP gap modulation with flux that corresponds to a shell coherence length $\xi = 140$~nm. (e) Wavefunction modulus ($|\Psi(r)|=|\tilde\Psi(r)|$) of the lowest excitation at $k_z=0$ in the normal state ($\Gamma_{\rm{N}}=0$) as a function of radial coordinate $r$ for a tubular-core semiconducting nanowire of external radius $R_{\rm{core}}=70$~nm and different thicknesses, from $d_{\rm{core}}=R_{\rm{core}}-R_{\rm{inner}}=10$~nm at the top to 70~nm (solid core) at the bottom. Different (normal) angular momentum subbands $m_l$ are indicated with different colors. (f) Same as (e) but for a solid-core semiconducting nanowire with a dome-like potential profile $U(r)$, see Eq. \eqref{pot}. Different values of $U_{\rm{max}}-U_{\rm{min}}$ from top to bottom are indicated (with $U_{\rm{max}}=0$ and exponent $\nu=2$). Other parameters like in Fig. \ref{fig:tubular}.}
   \label{fig:WF}
\end{figure*}

\subsection{Degeneracy point shifts and skewness}

The evolution of the subgap Van Hove singularities as we go from the hollow-core to the tubular-core nanowire of decreasing $R_\mathrm{inner}$ is shown in Fig. \ref{fig:tubular}, all the way to $R_\mathrm{inner}=0$. Note that in this model the electrostatic potential across the tubular core is kept uniform for simplicity. The most immediate effect of gradually reducing $R_\mathrm{inner}$ is a shift of the degeneracy points (originally located at the center of each LP lobe in the hollow-core nanowire) towards higher fields for $n>0$, see e.g. the difference between Fig.~\ref{fig:tubular}(b) and Fig.~\ref{fig:tubular}(c). The shift can cause the degeneracy points, together with the diamond-shaped gap below them, to exit the $n\neq 0$ lobes altogether, see Figs. \ref{fig:tubular}(e-g). In the process, the CdGM analogs become skewed towards higher fields relative to the center of the LP lobes, see Figs. \ref{fig:tubular}(d-f). For sufficiently small $R_\mathrm{inner}$, however, the skewness is inverted, see Figs. \ref{fig:tubular}(g-i). This inversion happens sooner at higher LP lobes. The skewness of CdGM analogs is thus found to be a direct consequence of the shift of the degeneracy points, which becomes the central concept in understanding the tubular-core nanowire.

The shift of degeneracy points can be readily understood in terms of the radial wavefunction profile of modes inside the core. In the hollow-core case we showed that degeneracy points appeared at integer normalized flux, as experienced by core states. The fact that this condition matched the integer normalized flux as experienced by the shell (center of LP lobes) was a consequence of the simplifying assumption that $R_\mathrm{core}=R_\mathrm{shell}=R_\mathrm{LP}$, since then the area spanned by the superconductor and core states coincided. Now, in the tubular-core model, as the core wavefunctions are allowed to spread inwards within the interval $r\in[R_\mathrm{inner}, R_\mathrm{core}]$, see Fig. \ref{fig:WF}(e), the flux they experience at a given magnetic field decreases with respect to the LP flux $\Phi$ through the shell. This shifts the degeneracy points towards higher magnetic fields.

An approximate analytical expression for the shift can be derived by considering that the flux experienced by the spread-out wavefunction is the same as if it were concentrated at its average radius $R_\mathrm{av} = \langle r\rangle$. The flux at which the degeneracy point happens in the $n=1$ lobe, $\Phi_\mathrm{dp}=\pi R_\mathrm{LP}^2B^{\rm{dp}}_z$, then becomes
\beq
\label{shift}
\Phi_\mathrm{dp}/\Phi_0 = (1+\delta n_\mathrm{dp}) = \frac{R_\mathrm{LP}^2}{R_\mathrm{av}^2}.
\eeq
We analyze the validity of this approximation in Fig.~\ref{fig:WF}.
Taking Fig. \ref{fig:tubular}(d) as a starting point, which corresponds to a tubular nanowire with $R_{\rm{inner}}=50$~nm and $R_{\rm{core}}=70$~nm, we show in Fig.~\ref{fig:WF}(a,b) the change in the dimensionless shift $\delta n_\mathrm{dp}$ for two different values of $R_{\rm{shell}}$ (and thus of $R_\mathrm{LP}/R_\mathrm{av}$). For clarity, we have fixed the shell gap $\Omega(\Phi)$ to a constant $\Omega(0)=0.23$~meV, so that the LP modulation does not obscure the degeneracy point shift. \editE{This is done taking $\xi\rightarrow 0$ in Eq. \eqref{LP3}, which results in a self energy given by \eqref{Eq:SELambda0}.} The corresponding LDOS with a flux-dependent $\Omega(\Phi)$ in the destructive regime is shown in panels Fig. \ref{fig:WF}(c,d) with $\xi = 140$~nm for comparison. \footnote{[Note that Figs. \ref{fig:tubular}(d) and \ref{fig:WF}(c) correspond to a wire with the same parameters except for the value of $\xi$, which puts the former in the non-destructive LP regime and the latter in the destructive one.]} It is clear from Eq. \eqref{shift} that for $R_\mathrm{av}<\sqrt{2/3}R_\mathrm{LP}$ we have $\delta n_\mathrm{dp}>1/2$, which pushes the degeneracy point out of the $n=1$ LP lobe. This is the case of Fig.~\ref{fig:WF}(b), where $R_\mathrm{LP}$ is increased by setting $R_\mathrm{shell}=90$~nm. In this situation, the $n=1$ degeneracy point does not correspond any more to a stable configuration for any value of magnetic field, since the fluxoid number in the ground state changes from $n=1$ to $n=2$ already at $\Phi/\Phi_0=3/2 < 1+\delta n_\mathrm{dp}$. Instead, the degeneracy point can be found in a metastable configuration of the $n>1$ LP lobe, represented inside the dashed white square of Fig.~\ref{fig:WF}(b). Note, however, that the skewness of the $n=1$ CdGM analogs does not depend on whether the degeneracy point is in a stable or metastable configuration.

In principle one needs to know the wavefunction profile to compute its $R_\mathrm{av}=\langle r\rangle$ in order to use Eq. \eqref{shift}. However, in the case of a uniform electrostatic potential, the wavefunction is approximately symmetric around the geometric mean radius $(R_\mathrm{inner}+R_\mathrm{core})/2$ of the core for all $m_l$, as long as $R_\mathrm{inner}/R_\mathrm{core}\gtrsim 0.5$, so that the effect of the Jacobian is small. This is shown in Fig. \ref{fig:WF}(e). Hence, we can use the approximation $R_\mathrm{av}\approx (R_\mathrm{inner}+R_\mathrm{core})/2$ in Eq. \eqref{shift}. This yields $\delta n_\mathrm{dp}=0.36$ for the parameters of Fig. \ref{fig:tubular}(a), which is very close the the numerical result of $0.40$ observed in that figure. The same happens for Fig. \ref{fig:tubular}(b), where the analytical solution is $\delta n_\mathrm{dp}=0.78$ and the numerical one is $0.83$. Therefore, we find that taking $R_\mathrm{av}$ as the average core radius and using it for the purpose of determining the flux that threads the wavefunction is a good approximation in the tubular-core model. Deviations are expected only when the wavefunction spreads substantially away from the core/shell interface. In this case different $m_l$ exhibit different $\langle r\rangle$, and the (metastable) degeneracy point becomes blurred and is no longer well defined. \editP{In Appendix  \ref{ap:dpoint} we discuss this effect in more detail}.

\begin{figure*}
   \centering
   \includegraphics[width=\textwidth]{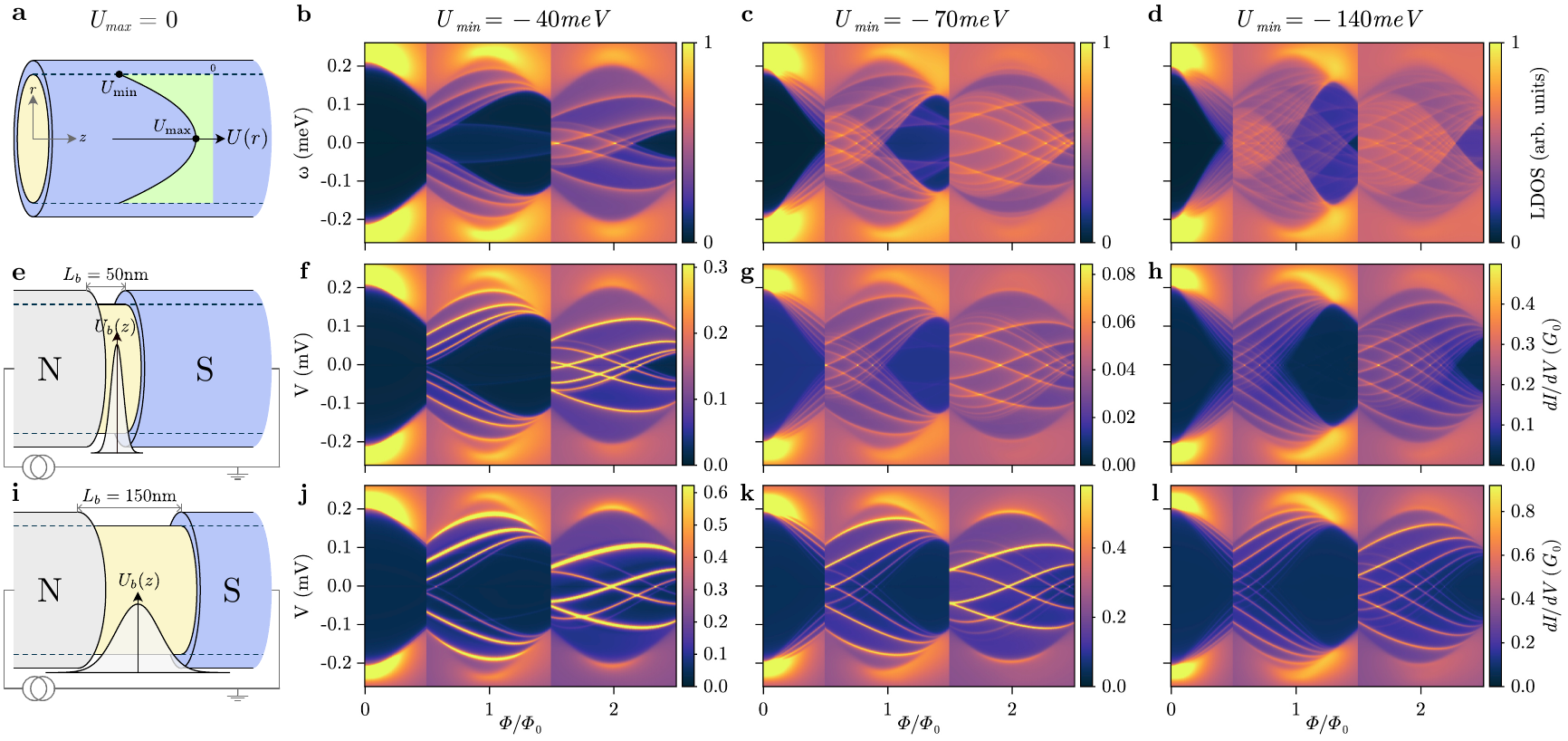}
   \caption{(a) Schematics of a solid-core, full-shell, semi-infinite nanowire with semiconductor electrostatic potential $U(r)$ in its interior. (b-c) LDOS in arbitrary units as a function of energy $\omega$ and normalized flux $\Phi/\Phi_0$ for a $R_{\rm{core}}=70$~nm, $R_{\rm{shell}}=80$~nm nanowire with shell coherence length $\xi = 70$~nm, core potential maximum $U_{\rm{max}}=0$ and different values of $U_{\rm{min}}$, from a shallow (b) to a deep (d) dome-like profile. (e) Schematics of a full-shell nanowire-based normal-superconductor tunnel junction. The potential-barrier profile $U_b(z)$ in the uncovered semiconductor region between the normal metal (N) and the full-shell wire (S) only depends on $z$. (f-h) Differential conductance $dI/dV$ (in units of the conductance quantum $G_0$) versus normalized flux for the same full-shell nanowires as in (b-d), and for a sharp tunnel barrier of width $50$~nm and height $60$~meV (f), $110$~meV (g) and $170$~meV (h). (i-l) Same as (e-l) but for a longer tunnel junction; width $150$~nm and heights $25$~meV (j), $50$~meV (k) and $110$~meV (l). Parameters: Column (b,f,j) has $\Gamma_{\rm{N}}=90\Omega(0)$; column (c,g,k) has $\Gamma_{\rm{N}}=30\Omega(0)$; and column (d,h,l) has $\Gamma_{\rm{N}}=20\Omega(0)$. Other parameters like in Fig. \ref{fig:tubular}.}
   \label{fig:solid}
\end{figure*}

\section{Solid-core nanowire}
\label{sec:solid}

\subsection{Model}

The development of the nanowire model culminates in this section, in which we consider a more accurate approximation to the actual full-shell wires studied in recent experiments ~\cite{Vaitiekenas:S20,Valentini:S21,Valentini:N22,Ibabe:A22}. These are all solid-core nanowires with $R_\mathrm{inner}=0$. We keep the cylindrical approximation, since we find that a more complicated hexagonal wire cross-section, which is computationally much more expensive, does not significantly affect the skewness and overall properties of the CdGM analogs (not shown here). We include, however, an extra crucial ingredient in the solid-core model: the non-homogeneous electrostatic potential $U(r)$ inside the core. This potential is a consequence of the band-bending imposed by the epitaxial core/shell Ohmic contact, which in turn stems from the difference of the Al work function and the InAs electron affinity \cite{Mikkelsen:PRX18}. We note that the degree of band bending and precise shape of $U(r)$ depend on the microscopic details of the interface and the self-consistent electrostatic screening. In keeping with our conceptual approach up to this point, we consider a simple model for $U(r)$ of the form
\beq
\label{pot}
U(r) = U_\mathrm{min} + (U_\mathrm{max}-U_\mathrm{min})\left(\frac{r}{R_\mathrm{core}}\right)^\nu,
\eeq
see illustrations in Fig. \ref{fig:sketch}(d) and Fig. \ref{fig:solid}(a). We specialize our simulations for $\nu=2$, $U_\mathrm{max}=0$, and $U_\mathrm{min}$ ranging from $-140$~meV to $-40$~meV, as suggested by microscopic calculations \cite{Mikkelsen:PRX18,Vaitiekenas:S20}. \editE{An equivalent analysis for $U_\mathrm{max}< 0$ is given in App. \ref{ap:Umax}.} The effective BdG Hamiltonian for the solid-core nanowire in the $m_L, k_z$-rotated basis then reads
\beqa
\label{solidrot}
\tilde H &=& \left[\frac{(m_L -\frac{1}{2}n\tau_z+\frac{1}{2}\frac{\Phi}{\Phi_0}\frac{r^2}{R_{\rm{LP}}^2} \tau_z)^2}{2m^*r^2}+ \frac{k_z^2+p_r^2}{2m^*} + U(r)\right]\tau_z \nonumber\\
&&+ \Sigma_\mathrm{shell}(\omega, 0).
\eeqa

\subsection{LDOS and transport}

Even though $R_\mathrm{inner}=0$ in the solid-core model, we find that the LDOS dependence with $\Phi$ for $R_\mathrm{core}=70$~nm, shown in Figs. \ref{fig:solid}(b-d) for three values of $U_{\rm{min}}$, is rather similar to that of tubular-core nanowires with $R_\mathrm{core}=70$~nm and $R_\mathrm{inner}=30-50$~nm, Figs.~\ref{fig:tubular}(d-f). The reason is that the potential $U(r)$ concentrates the wavefunction of the various modes to a region close to the core/shell interface. Indeed, we find that the wavefunction profiles for the different $m_l$ normal modes, depicted in Fig.~\ref{fig:WF}(f) for various values of $U_\mathrm{min}$, are very similar to those found for the tubular-core wire with varying semiconducting tube thicknesses, Fig. \ref{fig:WF}(e). Furthermore, the shape of the wavefunction in the solid-core model depends only weakly on the potential $U(r)$ for realistic values $U_\mathrm{min} < -40$~meV. The main effect of increasing the band bending (taking more negative $U_\mathrm{min}$ values) is increasing the number of occupied $m_l$ modes, and hence the number of CdGM analogs visible within each lobe [compare Fig. \ref{fig:solid}(b) to Fig. \ref{fig:solid}(d)].

\begin{figure}
   \centering
   \includegraphics[width=1\columnwidth]{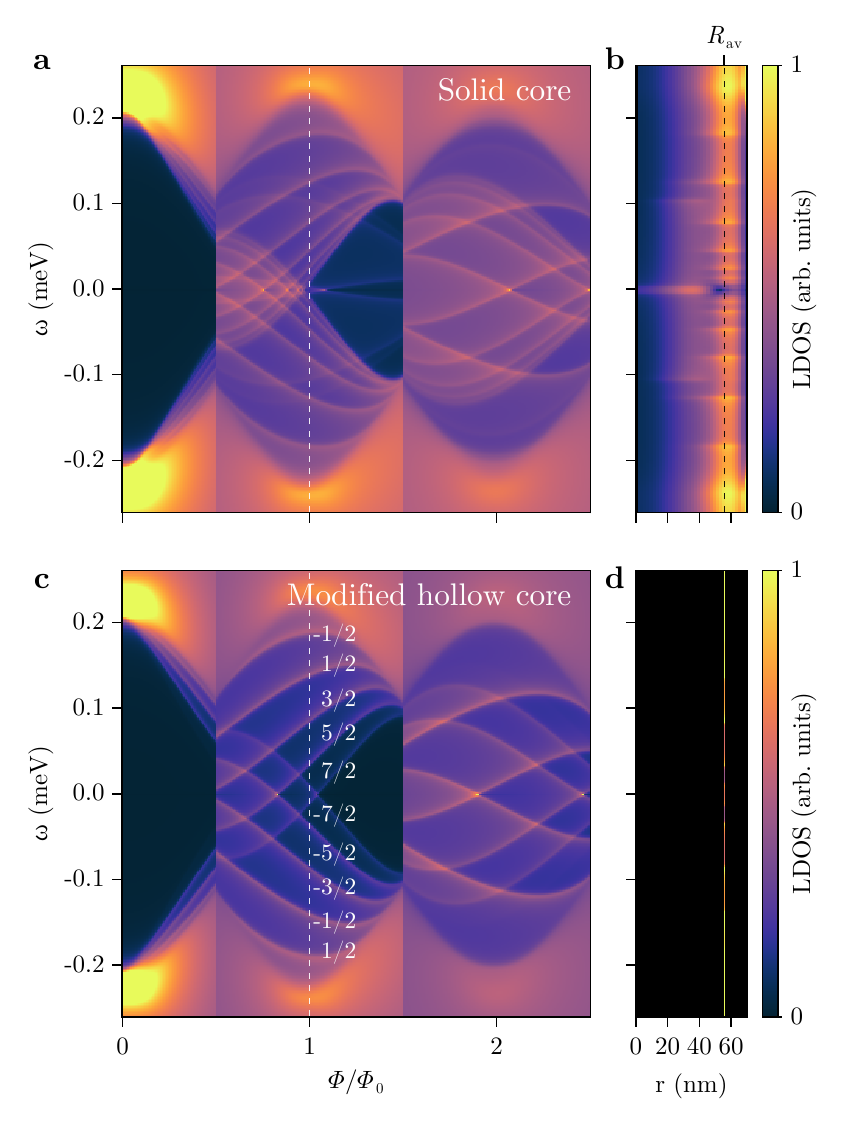}
   \caption{(a) Nanowire similar to that of Fig. \ref{fig:solid}(c), but with $a_0=2$~nm and $\Gamma_{\rm{N}} = 600\,\Delta(0)$. (b) Square root of the LDOS at the end of the nanowire as a function of energy $\omega$ and radial coordinate $r$ for flux $\Phi=\Phi_0$ (white dashed line) in (a). The black dashed line in (b) corresponds to the average radius $R_{\rm{av}} \approx 56$~nm. (c, d) Same as (a, b) for a modified hollow-core model with the same form of the shell-induced self-energy as in (a, b) but with $\Gamma_N=4.3\,\Delta(0)$. The solid core is replaced by a hollow cylinder of radius $R_{\rm av}$, with filling $\mu$ adjusted to keep the same number of occupied subbands (here $\mu=12.7$~meV with $a_0=2$~nm). Fractions in white denote $m_L$ for CdGM analogs in the first lobe. Note that the CdGM spectrum of the modified hollow-core model matches qualitatively that of the solid-core model, while being much more efficient to compute.}
   \label{fig:effective}
\end{figure}

In a typical experiment, the local subgap spectral structure in hybrid wires is measured via tunneling-transport spectroscopy. The technique measures the differential conductance $dI/dV$ between a normal probe at bias $V$ and a grounded wire, across a gate-tunable barrier. For high and sharp barriers, it was shown that the $dI/dV$ becomes proportional to the BdG LDOS at the contact~\cite{Bardeen:PRL61,Melo:SP21}, hence the name tunneling spectroscopy. We confirm this by computing $dI/dV$ in the non-interacting Green's function formalism across a sharp $L_b=50$~nm-long Gaussian $U_b(z)$ barrier, with a normal probe defined using the same model as the solid-core nanowire but without the shell-induced $\Sigma_\mathrm{shell}$. The resulting $dI/dV$ indeed matches the LDOS closely, see Fig. \ref{fig:solid}(f-h), with three notable differences that can be attributed to the small but finite $L_b$. First, Van Hove singularities appear much sharper in the $dI/dV$ than in the LDOS, resembling discrete subgap levels. Second, a small particle-hole $V\to-V$ asymmetry is visible in $dI/dV$, whereas the BdG LDOS is symmetric by definition. Last, and most significant, the small but finite $L_b$ makes transport more sensitive to modes with smaller $|m_L|$ values [see $m_L$ labels in Fig. \ref{fig:effective}(c), showing a case similar to Fig. \ref{fig:solid}(c)]. For smaller (larger) values of $|m_L|$, different subbands are deeper (shallower) in $k_z$-space, which translates into a slower (faster) evanescent decay inside the barrier. This makes the transmission probability through the barrier acquire a strong $m_L$ dependence as barrier length increases.\cite{Bardeen:PRL61}
As a result, Van Hove singularities with larger $|m_L|$ appear much fainter in the $dI/dV$, or they may even become undetectable.


Finally, Fig. \ref{fig:effective}(b) shows the radially-resolved LDOS in the middle of the $n=1$ lobe. This is closely related to the profile of the core wavefunctions that were shown in Fig. \ref{fig:WF}(f), but this time including their coupling to the shell. We confirm that most of the subgap states remain concentrated around a certain $R_\mathrm{av}$ inside the core [see dashed line in panel \ref{fig:effective}(b)]. We can see that, except for states close to the gap edge, there is a rather small leakage of core states into the shell, even for a strong proximity effect ($\Omega^*(0)\approx\Omega(0)$, perfect epitaxial contact). Notice also the existence of other fainter modes, particularly at low energy, with maxima away from $R_\mathrm{av}$ and closer to the nanowire axis. These correspond van Hove singularities in higher radial-momentum subbands, which may become populated in the low $|m_L|$ sectors for dense enough and/or thick enough nanowires.

The concentration of all the lowest radial subbands around a common $R_\mathrm{av}$ supports our interpretation of the consistent skewness of the CdGM analogs with flux. It also points to an interesting simplification of our solid-core model, which we dub the \emph{modified} hollow-core model, identical to the original hollow-core model in Sec. \ref{sec:hollow}, but with $R_\mathrm{core}$ replaced by $R_\mathrm{av}$. The $\mathcal{U}$-transformed $\tilde H$ thus reads
\beqa
\label{modifiedhollow}
\tilde H &=& \left[\frac{(m_L -\frac{1}{2}n\tau_z+\frac{1}{2}\frac{\Phi}{\Phi_0}\frac{R_{\rm av}^2}{R_{\rm{LP}}^2} \tau_z)^2}{2m^*R_{\rm{av}}^2}+ \frac{k_z^2}{2m^*} - \mu\right]\tau_z \nonumber\\
&&+ \Sigma_\mathrm{shell}(\omega, 0).
\eeqa
The self-energy $\Sigma_\mathrm{shell}$ is kept the same as in the solid core model, with the same dependence on $d_\mathrm{shell}$ and $R_\mathrm{LP}$, see Eq. \ref{LP3}. The value of $\Gamma_N$ in Eq. \eqref{Eq:SEmain}, however, needs to be adjusted to keep $\Omega^*(0)$ as in the solid core case. The definition of $\Phi$ and the LP lobes remain unchanged. The factor $R_{\rm{av}}^2/R_{\rm{LP}}^2$ in the angular kinetic energy produces the required shift $\delta n_{\rm dp}$ of the degeneracy points. The resulting modified hollow-core model above is almost trivial to solve numerically when compared to the solid core, and exhibits a similar phenomenology, as shown in the comparison of Fig. \ref{fig:effective}. 

\editE{It should be noted, nevertheless, that the modified hollow-core model is valid as long as the degeneracy point is well defined. In this case, all CdGM analogs from different $m_L$ sectors have approximately the same $R_{\rm{av}}$. When that is not the case, see App. \ref{ap:dpoint} for a discussion, this simple Hamiltonian ceases to be valid and a calculation with the full solid-core model is necessary.}

\section{Role of spin-orbit coupling and Zeeman splitting}
\label{sec:SOC}

Up to this point, we have neglected SOC and Zeeman splitting in our models. In the context of topological superconductivity, SOC is a crucial ingredient as it controls the magnitude of the topological gap and localizes the Majorana zero modes in these systems. However, as we show in this section, the SOC does not have a significant effect on the rest of the subgap spectrum, whereas the Zeeman coupling produces only a small splitting of otherwise degenerate CdGM analog states.

The generalization of the Hamiltonian in Eq. \eqref{solidrot} to include SOC was discussed in Ref. \onlinecite{Vaitiekenas:S20}. It involves writing the canonical transformation $\mathcal{U}$ in terms of the eigenvalues $m_J$ of the total generalized angular momentum, $J_z = -i\partial_\varphi +\frac{1}{2}\sigma_z+\frac{1}{2}n\tau_z$ (instead of the $m_L$ eigenvalues of the generalized orbital momentum $L_z$ used so far). In the presence of a radial SOC, the $m_J$'s are good quantum numbers of the Hamiltonian eigenstates even with Zeeman. Note that now $m_J$ is a half-integer (an integer) in the even (odd) lobes. The Zeeman effect can be included in the BdG Hamiltonian $\tilde H$ by adding the term
\beq
V_Z=\frac{1}{2}g\mu_BB_z\sigma_z,
\eeq
where $\mu_B$ is the Bohr magneton and $g$ is the nanowire Landé $g$ factor. Following this scheme, we computed the LDOS of a solid-core nanowire with and without Zeeman and SOC. We use $g=12$ for our simulations and a SOC of the form $\alpha(r)\hat{\bm{r}}\cdot(\bm{\sigma}\times \bm{p})$. Here $\alpha(r)$ is written, using a standard approximation from the 8-band model \cite{Winkler:03}, in terms of the radial potential gradient $\partial_r U(r)$,
\begin{equation}
\alpha(r) = \alpha_0\partial_r U(r) = \frac{P^2}{3}\left[\frac{1}{\Delta_g^2}-\frac{1}{(\Delta_\mathrm{soff}+\Delta_g)^2}\right]\partial_rU(r).
\label{Eq:SOC}
\end{equation}
Using the Kane parameter $P = 919.7~\mathrm{meV}\, \mathrm{nm}$, the semiconductor gap $\Delta_g = 417$~meV and split-off gap $\Delta_s=390$~meV, relevant for InAs, we obtain $\alpha_0 = 1.19~\textrm{nm}^2$ (see Ref. \onlinecite{Escribano:PRR20} for more elaborate approximations).

\begin{figure}
   \centering
   \includegraphics[width=1\columnwidth]{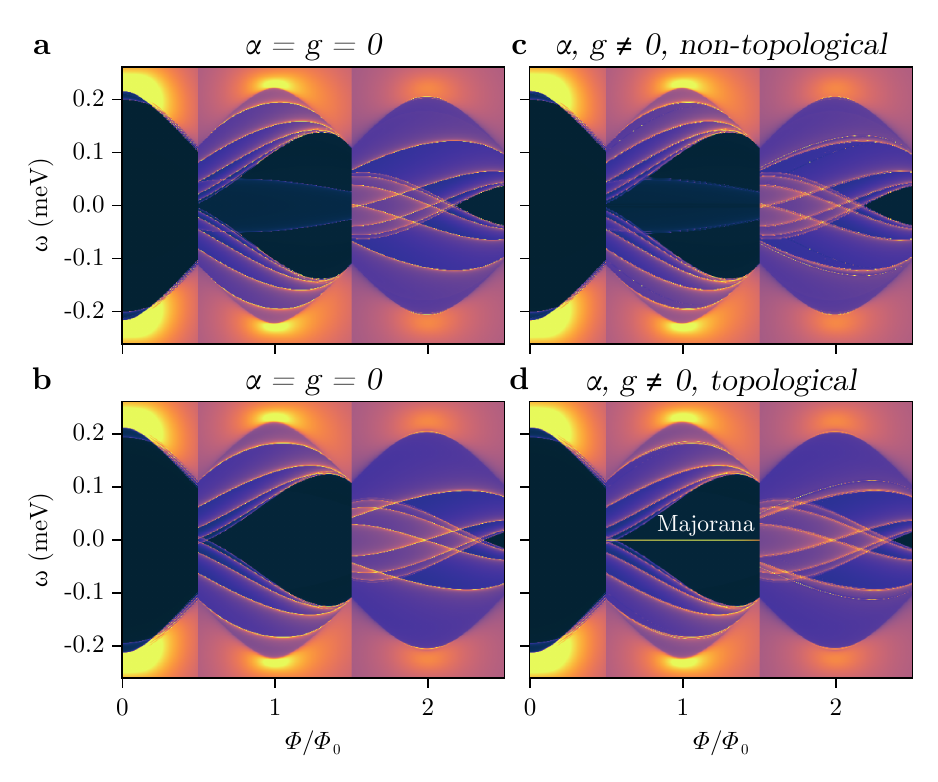}
   \caption{(a,b) LDOS at the end of a semi-infinite nanowire in arbitrary units versus energy $\omega$ and normalized flux $\Phi/\Phi_0$ at two different values of $U_\mathrm{min}=-40$~meV and $U_\mathrm{min}=-35$~meV, respectively, with no SOC ($\alpha$) or Zeeman ($g$). (c,d) The same configuration with a finite SOC and Zeeman coupling (see Sec. \ref{sec:SOC}), which correspond to a non-topological phase (c) and a topological phase with a Majorana zero mode (d), respectively. All other parameters like the first row of Fig. \ref{fig:solid}.}
   \label{fig:SOC}
\end{figure}

The simulations are shown in Figs. \ref{fig:SOC}(a,b) for $g=\alpha=0$ and in Figs. \ref{fig:SOC}(c,d) with finite $g$ and $\alpha$. In the two rows we choose  different values of $U_\mathrm{min}$ that result in topologically trivial and non-trivial phases for the $m_J=0$ sector when SOC is included, see Figs. \ref{fig:SOC}(c) and \ref{fig:SOC}(d), respectively. Note that the topological phase diagram of full-shell wires depends on $U_\mathrm{min}$ in a quite complex way, see for instance Fig. 4E of Ref. \onlinecite{Vaitiekenas:S20}. Thus, for some value of $U_\mathrm{min}$ the wire may be topological, while for a relatively similar one it may become trivial. In experimental hybrid nanowires, $U_\mathrm{min}$ depends sensitively on the microscopic details of the interface and can vary substantially depending on the growth conditions \cite{Schuwalow:AS21}. In the topological case of Fig. \ref{fig:SOC}(d) the SOC indeed produces a sharp Majorana zero mode with $m_J=0$ throughout the first lobe, as expected (recall that the nanowire is semi-infinite). More generally, SOC also introduces a few additional, very sharp subgap features, while Zeeman produces a small splitting of each Van Hove peak. Overall, however, the CdGM analog spectrum remains almost unperturbed. \editE{We have checked that this is also the case for larger values of $|U_\mathrm{min}|$, even though the magnitude of the SOC is larger according to Eq. \eqref{Eq:SOC}. We emphasize that the presence of SOC has profound consequences on this system, in the sense that it can drive the full-shell nanowire into the topological regime and trigger the appearance of Majorana bound states, but it otherwise leaves the rest of the subgap spectrum, i.e., the CdGM analog states, remarkably unaffected.}

\section{Summary and conclusions}
\label{sec:discussion}

We have shown that full-shell nanowires can host analogs of the CdGM states in type-II superconductors, which result from fluxoid quantization in the LP effect. Although the flux itself is not quantized as in an Abrikosov vortex, the integer number of superconducting-phase twists in finite-$n$ LP lobes similarly stabilizes a variety of low-energy subgap states in the nanowire core. Despite the various differences mentioned in the introduction between the original CdGM states and their full-shell analogs, their essential nature is the same: they are subgap states resulting from the two-dimensional confinement of a superconductor with a finite winding in its phase.

At a more detailed level, however, full-shell CdGM analogs are Van Hove singularities in propagating core subbands with a far richer structure. Their ring-like wavefunction within the wire cross section is maximal around a characteristic radius $R_\mathrm{av}$ that depends on the dome-like potential profile within the semiconductor. This turns out to have important consequences for their evolution with flux, $\Phi$. 

We found that degeneracy points appear, where all Van Hove singularities coalesce, and which shift toward larger absolute values of the threading flux $\Phi$ for $n\neq 0$ lobes, even disappearing beyond the lobe edge. This shift occurs as $R_{\rm{av}}$ is reduced relative to $R_{\rm{core}}$, or as $R_{\rm{LP}}$ increases with respect to $R_{\rm{core}}$.
The shift leaves behind a bundle of CdGM analogs that fill the whole LP-modulated parent gap $\Omega(\Phi)$ towards the left (right) side of each $n>0$ ($n<0$) lobe, but that tends to leave a characteristically shifted induced gap $\Omega^*(\Phi)$ on the opposite side. The degeneracy-point shift is proportional to $n$, so that the shifted gap is more visible in the $|n|=1$ lobe and tends to disappear faster with decreasing $R_{\rm{av}}$ for $|n|\ge 2$.

Furthermore, due to their orbital coupling to the field, CdGM analogs disperse with $\Phi$ with a positive or negative slope depending on their angular momentum $m_L$. For $\omega>0$, the states with negative $m_L$ are pushed toward the parent gap edge, leaving mostly $m_L>0$ states below the LP gap edge that exhibit a systematic skewness towards higher $\Phi$ values within finite-$n$ LP lobes, see e.g. Fig. \ref{fig:effective}. The $m_L$ quantum numbers of these $\omega>0$ CdGM analogs are ordered from smaller to larger values from the LP shell gap edge towards zero energy, while the opposite is true for $\omega<0$ states. By contrast, this is the opposite order than CdGM states in type-II vortex cores.

The number, energy, and skewness of the CdGM analogs in full-shell wires can be accessed experimentally using tunneling spectroscopy with sufficiently short tunnel barriers. The measured CdGM spectrum can be used to extract a wealth of otherwise inaccessible microscopic information about the electronic structure of the encapsulated nanowire. This includes details about the electrostatic potential profile inside the core, the resulting carrier density, its spatial distribution characterized by $R_\mathrm{av}$, the angular momentum of each mode or the transparency of the core/shell interface. The value of $R_\mathrm{av}$ can furthermore be used to define a modified hollow-core model that qualitatively captures most of the spectral features of the more complex microscopic model. Our analysis remains robust when introducing additional complexity, such as SOC, Zeeman splittings or even non-cylindrical nanowire cross sections.

The CdGM analogs could be measured using tunneling spectroscopy across a barrier at the end of the hybrid nanowire. In devices with longer tunneling barriers, we find that conductance becomes less sensitive to CdGM analogs with higher angular momentum. In experiments, subgap states localized at the barrier are sometimes observed. When present, they are not part of the bulk nanowire spectrum and should strongly depend on the barrier details. It should be possible to verify experimentally that a measured subgap feature in the tunneling conductance is indeed a CdGM analog, and not a barrier-localized state, by checking that its energy is largely insensitive to changes in the tunnel barrier. The CdGM analogs, as Van Hove singularities of the nanowire bandstructure, are associated to the extended bulk of the nanowire, not to the local barrier details. Two characteristic properties of the CdGM analogs are their skewness with flux and the presence of a shifted gap. Since CdGM analogs are in fact Van Hove singularities, they are only fully developed in sufficiently long nanowires. However, they should also be visible in shorter nanowires of length $L$ as a collection of longitudinally quantized levels with a level spacing $\sim 1/L$, concentrated around the energy of the $L\to\infty$ singularities \cite{Valentini:N22}.

\editP{All the code used in this manuscript is available at Zenodo \cite{zenodo}.}

\acknowledgments

This research was supported by Grants PGC2018-097018-B-I00, PID2021-122769NB-I00 and PID2021-125343NB-I00 funded by MCIN/AEI/10.13039/501100011033 and by ``ERDF A way of making Europe", the Comunidad de Madrid through Grant No. S2018/NMT-4511 (NMAT2D-CM), the Danish National Research Foundation, and a research grant (Project 43951) from Villum Fonden.

\appendix


\section{Little-Parks effect of a diffusive shell}
\label{ap:LP}

In this Appendix we summarize standard results for the dependence of the pairing amplitude $\Delta(\Phi)$ and superconductor energy gap $\Omega(\Phi)$ with flux $\Phi$ in a diffusive superconducting shell. We follow closely the presentation of Ref. \onlinecite{Ibabe:A22}.


The Abrikosov-Gor'kov theory \cite{Abrikosov:SPJ61,Skalski:PR64} describes a superconductor in the presence of \editE{a uniform concentration of} paramagnetic impurities. When the mean free path is small enough so that the superconductor is in the diffusive regime, the quasi-classical retarded Green's function is given by
\beqa
g^S(\omega)=\pi\nu_F\frac{\tau_x-u(\omega)\tau_0}{\sqrt{1-u(\omega)^2}},
\label{Eq:g}
\eeqa
where $\nu_F$ is the density of states at the Fermi level (in the normal state) and $\tau_i$ are Pauli matrices in Nambu space. The complex function $u(\omega)$ is obtained as the solution of
\beqa
u(\omega)=\frac{\omega}{\Delta(\Lambda)}+\frac{\Lambda}{\Delta(\Lambda)}\frac{u(\omega)}{\sqrt{1-u(\omega)^2}}.
\label{Eq:u}
\eeqa
It depends on the depairing parameter $\Lambda$ introduced by the spin-polarized impurities and on the pairing amplitude $\Delta(\Lambda)$, which at zero temperature is given by
\beqa
\Delta(\Lambda)=\frac{\nu_F V_{eph}}{2}\int_{-\omega_D}^{\omega_D}d\omega{\textrm{Re}}\frac{1}{\sqrt{u(\omega)^2-1}},
\eeqa
where Re is the real part, $V_{eph}$ is the phonon-mediated effective electron-electron interaction and $\omega_D$ the Debye frequency. These quantities are related by the BCS relation $\Delta(0)=2\omega_De^{-1/\nu_F V_{eph}}$. For a given $\Delta(\Lambda)$, Eq. \eqref{Eq:u} can be expressed as a fourth-order polynomial with root $u(\omega)$ chosen so as to satisfy the appropriate continuity and asymptotic behaviors for retarded Green's functions. \editP{This implies, in particular, $u(\omega\to 0) = 0$}.

For finite pair-breaking, Abrikosov-Gor'kov found a closed form solution for the pairing amplitude
\beqa
\ln\frac{\Delta(\Lambda)}{\Delta(0)} &=& -P\left(\frac{\Lambda}{\Delta(\Lambda)}\right),\nonumber\\
P(z\leq 1) &=& \frac{\pi}{4}z,\nonumber\\
P(z\geq 1) &=& \ln\left(z+\sqrt{z^2-1}\right)+\frac{z}{2}\arctan\frac{1}{\sqrt{z^2-1}}\nonumber\\
&&-\frac{\sqrt{z^2-1}}{2z},
\label{LP1}
\eeqa
where $\Delta(0)$ is the pairing of the pure (ballistic) superconductor, i.e., for $\Lambda=0$. Note that $\Lambda$ has energy units and is bounded by $0\leq \Lambda\leq \Delta(0)/2$. The equation for $\Delta(\Lambda)$ has to be solved self-consistently.

Subsequently, Skalski \textit{et al.}\cite{Skalski:PR64} found an analytical expression for the energy gap, defined by the edge of the branch cut at $u(\Omega)^2 = 1$ in Eq. \eqref{Eq:g}, and given by
\beqa
\Omega(\Lambda) =  \left(\Delta(\Lambda)^{2/3}-\Lambda^{2/3}\right)^{3/2}.
\label{LP2}
\eeqa
Note that the energy gap $\Omega$ is only equal to the pairing amplitude $\Delta$ in the absence of depairing effects, and is smaller otherwise. There even exists a region of $\Lambda$ close to $\Delta(0)/2$ for which the gap in the excitation spectrum is zero even though the shell is still a superconductor in the sense of having a non-zero order parameter. This is the regime of so-called gapless superconductivity.

The problem of a superconductor containing paramagnetic impurities is very similar to the problem of an ordinary diffusive superconductor in the presence of an external magnetic field \cite{Maki:PTP64,Groff:PR68}. Thus, we can identify the depairing parameter produced by the magnetic impurities above with an analogous depairing produced by the magnetic flux, $\Lambda(\Phi)$. Assuming now cylindrical symmetry, a standard Ginzburg-Landau theory of the LP effect \cite{Lopatin:PRL05,Shah:PRB07,Dao:PRB09,Schwiete:PRB10,Sternfeld:PRL11} provides an explicit connection between flux and depairing
\beqa
\Lambda(\Phi) &=& \frac{\xi^2 k_B T_c}{\pi R_{\rm{LP}}^2}\left[4\left(n-\frac{\Phi}{\Phi_0}\right)^2 + \frac{d_{\rm{shell}}^2}{R_{\rm{LP}}^2}\left(\frac{\Phi^2}{\Phi_0^2} + \frac{n^2}{3}\right)\right],\nonumber\\
n(\Phi) &=& \lfloor \Phi/\Phi_0\rceil = 0, \pm 1,\pm 2, \dots,
\label{LP3}
\eeqa
where $\xi$ is the diffusive superconducting coherence length and $T_c$ is the zero-flux critical temperature. At zero field $\Lambda(0)=0$, $\Omega(0) = \Delta(0)$ and $k_B T_c\approx\Omega(0)/1.76$, where $k_B$ is the Boltzmann constant. 

The solution for Eqs. \eqref{LP1}-\eqref{LP3} is qualitatively different depending on the ratios $R_\mathrm{LP}/\xi$ and $d_\mathrm{shell}/R_\mathrm{LP}$. It ranges from the non-destructive regime ($\Omega$ is non-zero, satisfied for $R_\mathrm{LP}/\xi\gtrsim 0.6$ if $d_\mathrm{shell}\to 0$) to the destructive regime ($\Omega$ vanishes in a finite window around odd half-integer $\Phi/\Phi_0$, satisfied for smaller $R_\mathrm{LP}/\xi$)\cite{Schwiete:PRL09}. The different regimes both for $\Delta$ and $\Omega$ are represented in Fig.~\ref{fig:LP}. As a guideline, some typical values representative of recent experiments \cite{Vaitiekenas:PRB20} are $\xi\sim 100$~nm, $R_\mathrm{core}\sim 70$~nm, $d_\mathrm{shell}\sim 10$~nm and $\lambda_L=150$~nm. These parameters correspond to a superconductor in the dirty limit, which is the regime where the above theory is applicable, and the one relevant to current experiments.

\begin{figure*}
   \centering
   \includegraphics[width=\textwidth]{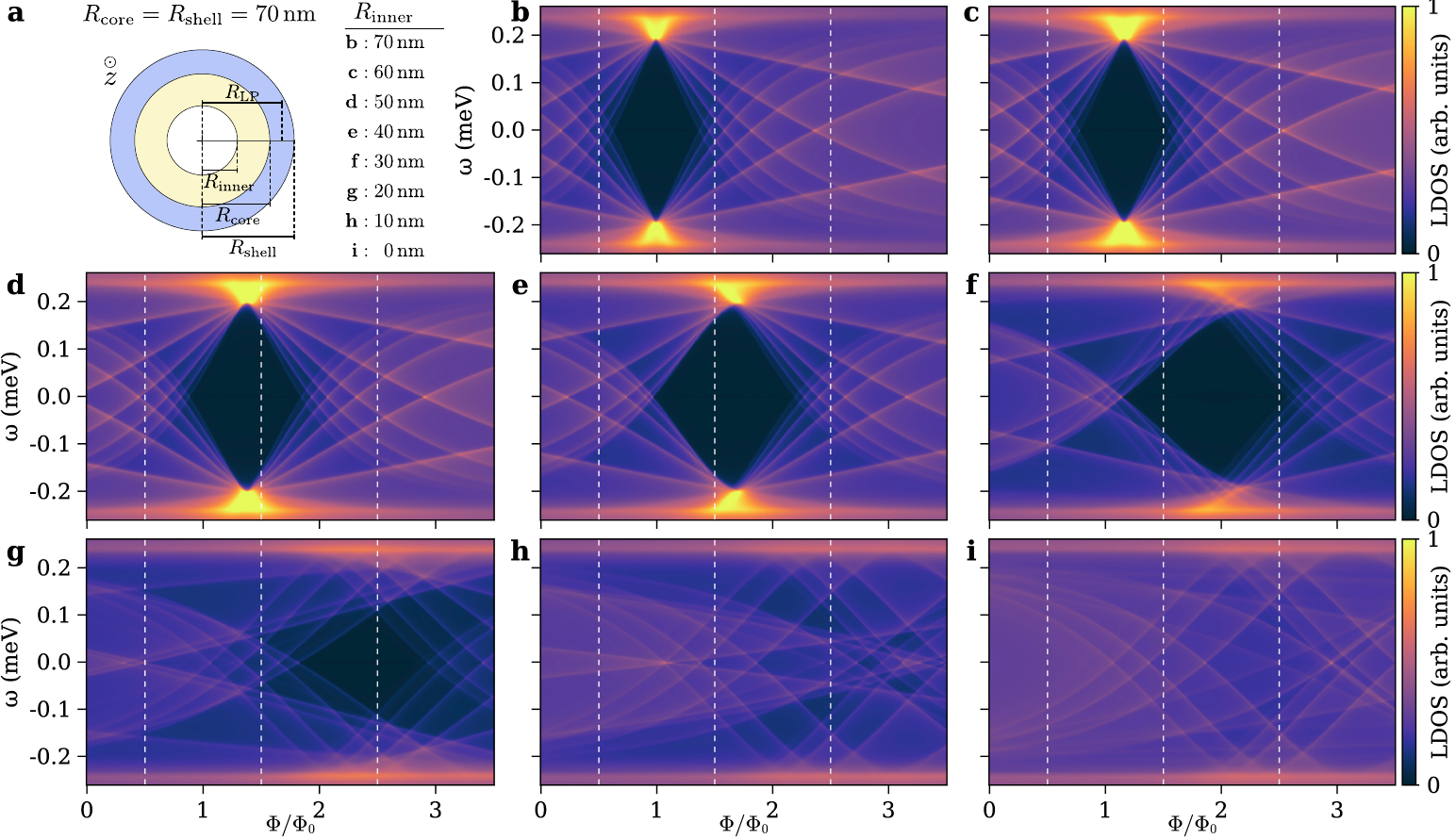}
   \caption{\editP{Evolution of the degeneracy point for decreasing $R_{\rm av}$ in the tubular core model. This figure is analogous to Fig. \ref{fig:tubular} save for three changes: the LP modulation is suppressed ($\xi\to 0$), the fluxoid number is forced to remain at $n=1$ for all $\Phi$, and $\Phi$ reaches up to the third lobe. The original lobes are delimited by dashed white lines. Decreasing $R_{\rm av}$ not only shifts the degeneracy point, but also makes it increasingly smeared, until it eventually dissolves.}}
   \label{fig:tubular_appendix}
\end{figure*}

\begin{figure*}
   \centering
   \includegraphics[width=\textwidth]{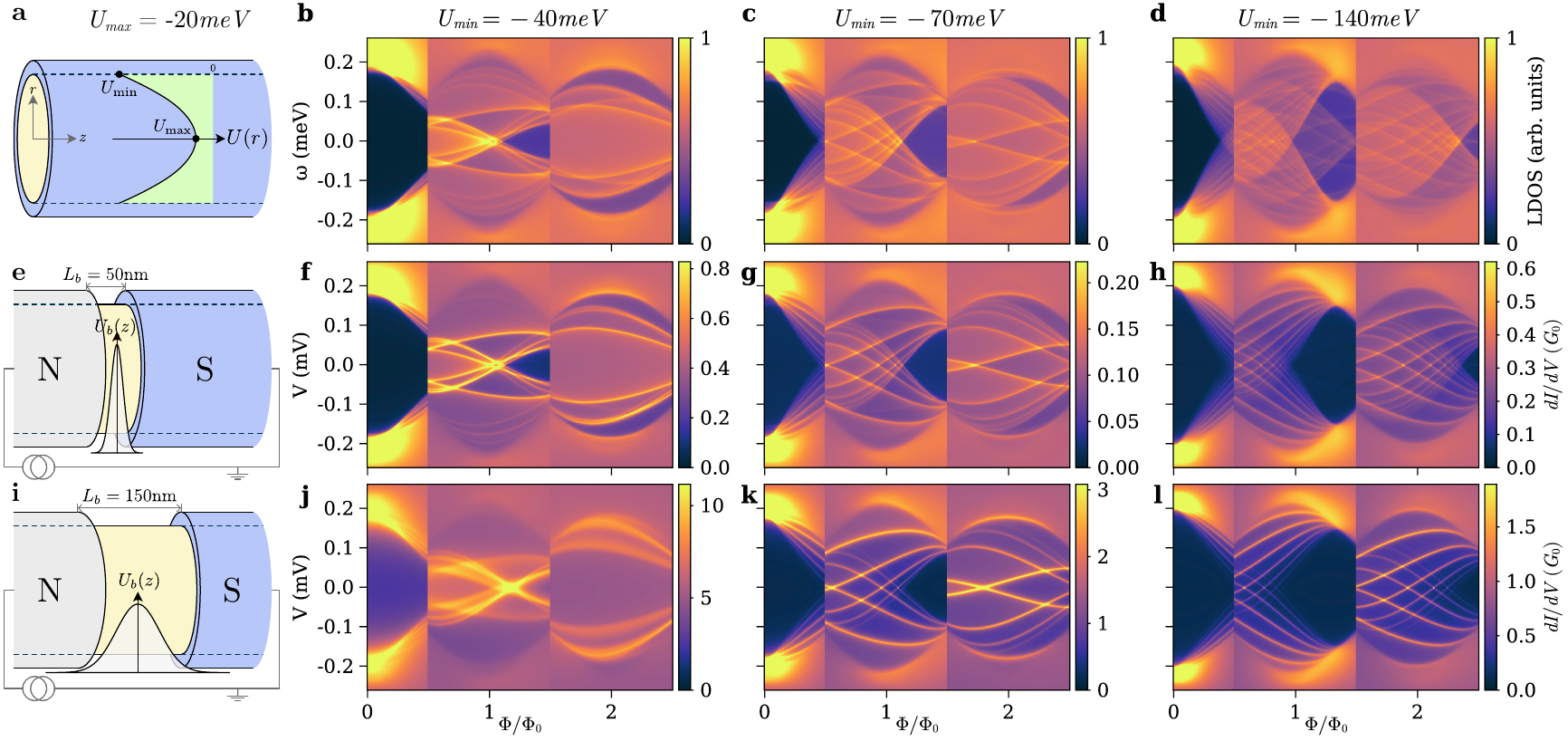}
   \caption{\editP{Effect of a finite $U_{\rm{max}}$ in LDOS and transport. This figure is identical to Fig. \ref{fig:solid}, save for the value of $U_{\rm{max}}$, which is here $-20$~meV instead of zero.}}
   \label{fig:solid_appendix}
\end{figure*}

\section{Self-energy from a diffusive shell}
\label{ap:SE}

The proximity effect of the diffusive superconducting shell described in App. \ref{ap:LP} onto the semiconducting core of the full-shell nanowire can be accounted for by means of a self-energy $\Sigma_\mathrm{shell}$ acting on the core's surface. Using a tight-binding language where $t_I$ is the hopping parameter between a surface site of the core lattice and the shell, the shell self-energy reads
\beq
\Sigma_\mathrm{shell}(\omega,\varphi)=t_I^2g^S(\omega,\varphi),
\eeq
where $g^S(\omega,\varphi)$ is the same one of Eq. \eqref{Eq:g} but including now the dependence of the pairing amplitude with the fluxoid number $n$ and the polar angle $\varphi$,
\beq
g^S(\omega,\varphi)=\pi\nu_F\frac{\editE{\cos(n\varphi)\tau_x+\sin(n\varphi)\tau_y-u(\omega)\tau_0}}{\sqrt{1-u(\omega)^2}},
\label{Eq:gphi}
\eeq
\editE{given in the Nambu basis $\Psi=(\psi^{\dagger}_{\uparrow},\psi^{\dagger}_{\downarrow},\psi_{\downarrow},-\psi_{\uparrow})$. }
Defining $\Gamma_{\textrm{N}}=\pi\nu_F^St_I^2$ as the normal decay rate from the core into the shell, we can write
\beq
\Sigma_\mathrm{shell}(\omega,\varphi)=\Gamma_{\textrm{N}}\frac{\editE{\cos(n\varphi)\tau_x-\sin(n\varphi)\tau_y}-u(\omega)\tau_0}{\sqrt{1-u(\omega)^2}},
\label{Eq:SE}
\eeq
where $u(\omega)$ is given in Eq. \eqref{Eq:u}.

\editE{In the canonical basis where the Hamiltonian transforms into a $\varphi$-independent effective Hamiltonian $\tilde H$, see Sec. \ref{subsec:quantnumb} of the main text, the self energy simplifies to
\beq
\Sigma_\mathrm{shell}(\omega,0)=\Gamma_{\textrm{N}}\frac{\tau_x-u(\omega)\tau_0}{\sqrt{1-u(\omega)^2}}.
\eeq
There are two interesting limits of this self energy. When $\omega\rightarrow 0$, then $u(\omega)\rightarrow 0$ and thus
\beq
\label{Eq:SEomega0}
\Sigma_\mathrm{shell}(0,0)=\Gamma_{\textrm{N}}\tau_x.
\eeq
This $\omega$-independent expression is correct when there is a tunnel coupling between the superconductor and the semiconductor. It has been used to calculate the band structure of Fig. \ref{fig:bands}.}

\editE{On the other hand, when the diffusive coherence length $\xi\rightarrow 0$ or, equivalently, when $\Lambda(\Phi)=0$ in Eq. \eqref{LP3}, then the LP modulation is disconnected, see Figs. \ref{fig:WF}(a,b) and \ref{fig:tubular_appendix}. In this case $u(\omega)=\omega/\Omega(0)$ in Eq. \eqref{Eq:u} and
\beq
\label{Eq:SELambda0}
\Sigma_\mathrm{shell}(\omega,0)=\Gamma_{\textrm{N}}\frac{\Omega(0)\tau_x-\omega\tau_0}{\sqrt{\Omega(0)^2-\omega^2}}.
\eeq
This expression corresponds to the conventional ballistic BCS self energy. It has been used in Fig. \ref{fig:WF}(a,b) and will be also used in the next Appendix.}

\section{Degeneracy point structure for decreasing $R_{\rm{av}}$}
\label{ap:dpoint}

\editP{We have seen in Sec. \ref{sec:tubular} that the degeneracy point of the first lobe (where all the CdGM analogs from different $m_L$ sectors converge) becomes shifted to higher flux as the average radius $R_{\rm av}$ decreases, eventually exiting the lobe. In this Appendix we demonstrate that there is an extra effect that arises for large shifts: the convergence of the CdGM analogs becomes increasingly imprecise, leading to a smearing of the degeneracy point and even to its complete dissolution at small enough $R_{\rm av}$.}

\editP{The reason for this is simple. As the wavefunction is allowed to spread throughout a larger range of radii away from the core/shell interface, the assumption that all $m_L$ have the same radial wavefunction profile, and hence share the same $R_{\rm av}$, is no longer a valid approximation, see Fig. \ref{fig:WF}(e,f). Since the $R_{\rm av}$ directly controls the flux at which the degeneracy point appears, each $m_L$ reaches the degeneracy condition (i.e., enclosing a flux quantum $\Phi_0$) at different magnetic field values, thus smearing the degeneracy point.}

\editP{The smearing effect is demonstrated in Fig. \ref{fig:tubular_appendix}, which is similar to Fig. \ref{fig:tubular}, but with the fluxoid number fixed to $n=1$ for all $\Phi$. To clearly follow the evolution of the degeneracy point we have also fixed the gap $\Omega$ to be $\Phi$-independent like in Fig. \ref{fig:WF}(a,b), and extended the flux range up to the third lobe. We see how, in addition to the shift to higher flux values, the degeneracy point becomes blurred for $R_{\rm{av}} \lesssim 55$~nm ($R_{\rm{inner}} \lesssim 40$~nm), and is completely dissolved for $R_{\rm{av}} \lesssim 45$~nm ($R_{\rm{inner}} \lesssim 20$~nm).}

\section{Solid-core nanowire results for $U_\mathrm{max}< 0$}
\label{ap:Umax}

\editE{For completeness, we explore in this Appendix the effect of reducing $U_{\rm max}$ in the solid core model to a finite negative value. This increases the density around the core axis. If we do not simultaneously decrease $U_{\rm min}$, this has the effect of moving the solid-core system closer to the tubular core model, which has a uniform $U(r)$. However, if $|U_{\rm min}|\gg|U_{\rm max}|$, the effect of the finite $U_{\rm max}$ is small. This is showcased in Fig. \ref{fig:solid_appendix}, which is identical to Fig. \ref{fig:solid}, but with $U_{max} = -20$~meV instead of 0~meV. We see that in the first column, the subgap spectrum becomes more similar to Figs. \ref{fig:tubular}(f-i), with states of negative skewness beginning to appear. However, for larger $|U_{\rm min}|$ (second and third columns), the change with respect to Fig. \ref{fig:solid} is minor.}

\bibliography{biblio}

\end{document}